\documentclass[12pt,hyperref]{article}
\usepackage{hyperref}
\usepackage{url,color}
\usepackage[final]{graphicx}  
\usepackage{caption}          
\usepackage{microtype}        
\usepackage[noabbrev]{cleveref} 
\usepackage{amsmath}
\usepackage{feynmp-auto}      
\usepackage{cancel, slashed}  
\usepackage{hepparticles}     
\usepackage{hepnames}         
\usepackage{subcaption}
\usepackage{diagbox}
\microtypesetup{
  activate={true,nocompatibility},  
  kerning=true, 
  spacing=true,
  protrusion=true
 }
\allowdisplaybreaks
\microtypecontext{spacing=nonfrench}
\makeatletter
\newif\if@preliminary
\@preliminaryfalse
\def\preliminary{\@preliminarytrue}

\def\preprintno#1{\def\@preprintno{#1}}
\def\address#1{\def\@address{#1}}
\def\email#1#2{\thanks{\tt #1@{}#2}}
\def\abstract#1{\def\@abstract{#1}}
\renewcommand\abstractname{ABSTRACT}
\newlength\preprintnoskip
\newlength\abstractwidth
\@titlepagetrue
\renewcommand\maketitle{\begin{titlepage}%
  \let\footnotesize\small
  \hfill\parbox{\preprintnoskip}{%
  \begin{flushright}\@preprintno\end{flushright}}\hspace*{1cm}
  \vskip 60\p@
  \begin{center}%
    {\Large\bf\boldmath \@title \par}\vskip 1cm%
    {\sc\@author \par}\vskip 3mm%
    {\@address \par}%
    \if@preliminary
      \vskip 2cm {\large\sf PRELIMINARY DRAFT \par \@date}%
    \fi
  \end{center}\par
  \@thanks
  \vfill
  \begin{center}%
    \parbox{\abstractwidth}{\centerline{\abstractname}%
    \vskip 3mm%
    \@abstract}
  \end{center}
  \end{titlepage}%
  \setcounter{footnote}{0}%
  \let\thanks\relax\let\maketitle\relax
  \gdef\@thanks{}\gdef\@author{}\gdef\@address{}%
  \gdef\@title{}\gdef\@abstract{}\gdef\@preprintno{}
}%
\def\@citex[#1]#2{\if@filesw\immediate\write\@auxout{\string\citation{#2}}\fi
  \def\@citea{}\@cite{\@for\@citeb:=#2\do
    {\@citea\def\@citea{,\penalty\@m}\@ifundefined
       {b@\@citeb}{{\bf ?}\@warning
       {Citation `\@citeb' on page \thepage \space undefined}}%
\hbox{\csname b@\@citeb\endcsname}}}{#1}}
\def\citerange{\@ifnextchar [{\@tempswatrue\@citexr}{\@tempswafalse\@citexr[]}}
\def\@citexr[#1]#2{\if@filesw\immediate\write\@auxout{\string\citation{#2}}\fi
  \def\@citea{}\@cite{\@for\@citeb:=#2\do
    {\@citea\def\@citea{--\penalty\@m}\@ifundefined
       {b@\@citeb}{{\bf ?}\@warning
       {Citation `\@citeb' on page \thepage \space undefined}}%
\hbox{\csname b@\@citeb\endcsname}}}{#1}}
\long\def\@makecaption#1#2{%
  \vskip\abovecaptionskip
  \sbox\@tempboxa{#1: \emph{#2}}%
  \ifdim \wd\@tempboxa >\hsize
    #1: \emph{#2}\par
  \else
    \hbox to\hsize{\hfil\box\@tempboxa\hfil}%
  \fi
  \vskip\belowcaptionskip}
\makeatother
\newcommand{\ii}{\mathrm{i}}

\newcommand{\LL}{\mathcal{L}}

\newcommand{\vW}{\mathbf{W}}
\newcommand{\vH}{\mathbf{H}}

\newcommand{\tr}[1]{\operatorname{tr}\left[#1\right]}
\newcommand{\vB}{\mathbf{B}}
\newcommand{\vD}{\mathbf{D}}

\newcommand{\GeV}{\text{GeV}}
\newcommand{\TeV}{\text{TeV}}

\newcommand{\whizard}{\texttt{WHIZARD}}
\newcolumntype{M}[1]{>{\centering\arraybackslash}m{#1}}
\newcolumntype{L}[1]{>{\arraybackslash}m{#1}}
\newcolumntype{N}{@{}m{0pt}@{}}
\makeatletter
\font\manfnt=manfnt
\def\Watchout{\@ifnextchar [{\W@tchout}{\W@tchout[1]}}
\def\W@tchout[#1]{{\manfnt\@tempcnta#1\relax%
  \@whilenum\@tempcnta>\z@\do{%
    \char"7F\hskip 0.3em\advance\@tempcnta\m@ne}}}
\def\remark{\@ifnextchar[{\@remark}{\@remark[1]}}

\def\@remark[#1]#2{%
  \setbox\@tempboxa\hbox{\W@tchout[#1]}
  \@tempdima\wd\@tempboxa
  \list{}%
    {\leftmargin\@tempdima}%
    \item[\hbox to 0pt{\hss\W@tchout[#1]}]%
    \textbf{[#2]}}
\makeatother

\begin{document}


\preprintno{%
  DESY 18-002\\
  SI-HEP-2018-17 \\
  KA-TP-14-2018 
} 

\title{%
  Transversal Modes and Higgs Bosons\\
  in Electroweak Vector-Boson Scattering\\
  at the LHC 
  }

\author{
  Simon Brass\email{simon.brass}{uni-siegen.de}$^{a}$,
  Christian Fleper\email{christian.fleper}{uni-siegen.de}$^a$,
  Wolfgang Kilian\email{kilian}{physik.uni-siegen.de}$^{a,d}$,
  J\"urgen Reuter\email{juergen.reuter}{desy.de}$^b$,
  Marco Sekulla\email{marco.sekulla}{kit.edu}$^c$
}

\address{\it%
    $^a$University of Siegen, Department of Physics,
    D--57068 Siegen, Germany, \\
    $^b$DESY Theory Group,
    D--22607 Hamburg, Germany,\\
    $^c$Institute for Theoretical Physics, Karlsruhe Institute of
    Technology, D--76128 Karlsruhe, Germany,\\
    $^d$CLICdp and Theory Group, CERN, CH-1211 Geneva 23,
    Switzerland
}

\date{%
  \today
}

\abstract{%
  Processes where $W$ and $Z$ bosons scatter into pairs of electroweak
  bosons $W$, $Z$, and Higgs, are sensitive probes of new physics in
  the electroweak sector.  We study simplified models that describe typical
  scenarios of new physics and parameterize the range of possible LHC
  results between the Standard-Model prediction and unitarity limits.
  Extending the study beyond purely longitudinal scattering, we
  investigate the role of transversally polarized gauge bosons.
  Unitarity becomes an essential factor, and
  limits on parameters matched to the naive perturbative low-energy
  effective theory turn out to be necessarily model-dependent.  We discuss the
  implications of our approach for the interpretation of LHC data on
  vector-boson scattering and Higgs-pair production.
}
\maketitle


\tableofcontents


\section{Introduction}
\label{sec:intro}

After the discovery of the Higgs boson~\cite{Aad:2012tfa,Chatrchyan:2012xdj},
particle physics faces the question whether the new scalar sector is minimal
or non-minimal, whether it is weakly or strongly interacting, and whether it
validates or invalidates the accepted paradigm of quantum field theory as a
universal description of particle interactions.  Experimental data on
electroweak boson interactions (Higgs, $W$, $Z$, and photon) will deepen our
understanding in that area.  Processes of the type $VV\to VV$, where $V=W,Z,H$
(vector-boson scattering, VBS, and associated or Higgs pair production in
vector-boson fusion, VBF), are a
most sensitive probe of electroweak physics and the Higgs sector.  They will
be extensively studied in the present and future runs of the LHC.

There is obvious interest in scenarios where new degrees of freedom
beyond the Standard Model (SM) couple primarily to the Higgs-Goldstone
boson field.  The presence of new physics coupled to the Higgs sector
might solve some of the long-standing puzzles of particle physics.
Such new modes need not have a significant effect on existing
precision data.  They might be strongly-interacting as in
composite Higgs models, or weakly interacting as in models with extra
scalars that are decoupled from SM fermions.  They should manifest
themselves primarily in interactions of massive electroweak bosons,
namely $W^\pm$, $Z$, and the Higgs itself.

The ATLAS and CMS experiments at the LHC have measured VBS processes
as a signal, embedded in partonic processes of the type $qq\to VVqq$,
where $q$ is any light quark.  Numerical results have been presented
in the form of limits on parameters within the SM effective theory
(SMEFT)~\cite{Buchmuller:1985jz,Hagiwara:1992eh,Hagiwara:1993ck,Grzadkowski:2010es}.
The usual application of the SMEFT truncates the power expansion of
the Lagrangian at the level of dimension-six operators.  A useful
parameterization of VBS processes requires dimension-eight
effective operators, the second order of the low-energy expansion
beyond the SM.

In recent work~\cite{Kilian:2014zja,Kilian:2015opv,Fleper:2016frz}, we
have studied deviations from the SM interactions that are confined to
the longitudinal scattering modes of $W$ and $Z$. In the low-energy
limit, the contributing set of new interactions is small, and if
custodial (weak isospin) symmetry is imposed, there are just two free
parameters in the matching SMEFT expansion.  On the other hand, recent
LHC analyses quote results for a larger set of operator coefficients
which include interactions between longitudinal and transversal modes
of $W$ and $Z$ bosons.  In the current paper, we study deviations from
the SM in VBS processes that involve transversally polarized $W$ and
$Z$ bosons, and also consider Higgs bosons in the final state.

Numerical results of non-SM interactions of longitudinal
scattering have clearly shown that for the level of deviations that
can be detected by the LHC experiments, the unitarity limits are
always violated in the high-energy range, if a naive SMEFT calculation
is attempted.  A model-independent parameterization beyond the SM that
covers the accessible parameter range becomes impossible.
Nevertheless, reasonable assumptions on new physics lead to unitarity
constraints that limit the level of possible excess above the SM
prediction.  With this knowledge, it is possible to devise simplified
models that both satisfy unitarity over the whole energy range and
smoothly match onto the SMEFT parameterization at low energy.  For the
purpose of an exemplary study, we have compared a class of
``continuum'' models which merely extrapolate the SMEFT expansion into
asymptotically strong interactions with models that describe single
resonances with specific quantum-number assignments.

In the present paper, we extend this program to also describe
transversal modes, and to include the Higgs boson as a possible final
state on the same footing as the $W$ and $Z$ bosons.  Regarding the SM
processes as reference, there are detailed
calculations~\cite{Jager:2009xx,Melia:2010bm,Melia:2011dw,Jager:2011ms,Baglio:2014uba,Biedermann:2016yds}
beyond leading-order in the SM perturbation theory. Recently, there
was a concise comparison of several different codes for the precision
simulation at LO and NLO for like-sign vector boson scattering at 13
TeV~\cite{Ballestrero:2018anz}.  For
the simplified models considered in this paper, we confine ourselves
to leading-order calculations but we remark that adding in
perturbative QCD and electroweak corrections is possible by the same
methods as for the pure SM, and should eventually be done in order to
distinguish genuine deviations from uncertainties of the
approximation, in actual data analysis.

The paper is organized as follows.  In Sec.~\ref{sec:currents} we
discuss the structure of new interactions in the electroweak and Higgs
sector, and state the underlying assumptions.  This defines the SMEFT
ansatz, and it allows us to list the operators that describe the
low-energy limit.  The symmetries of those interactions determine the
eigenmodes of quasi-elastic $2\to 2$ scattering, which allows us to
diagonalize the amplitudes for all vector-boson modes,
Sec.~\ref{sec:diagonal}.  We construct unitary
models exhibiting a strongly interacting continuum in
Sec.~\ref{sec:contmodel}.  These models would yield the maximally allowed
number of events consistent with quantum field theory in the VBS
channel, matched to the low-energy SMEFT with specific values for the operator
coefficients.  In Sec.~\ref{sec:sim_res} we discuss
simplified models which contain a resonance and likewise parameterize VBS
amplitudes at all energies.  We present numerical results and plots
for selected parameter sets and final states.  In Sec.~\ref{sec:results},
we discuss the relevance of our study in view of
future analyses at the LHC and beyond.

\section{Electroweak Boson Currents and Local Operators}
\label{sec:currents}

Expectations for new physics beyond the SM are constrained by available data.
They may also be guided by imposing principles such as simplicity and absence
of accidental cancellations.  For the current work, we base the description of
new effects beyond the SM on the following assumptions: (i) light fermions do
not participate directly in new dynamics, (ii) the observed pattern of SM
gauge invariance retains its relevance beyond the $\TeV$ range, and (iii) new
degrees of freedom beyond the SM do not carry open color.  These assumptions
are not mandatory but backed by the available precision data regarding flavor
physics, QCD, new-physics searches at the LHC, and precision electroweak
observables.

If these assumptions are accepted as guiding principles for a phenomenological
description, a parameterization of dominant new effects can qualify
light-fermion currents as classical spectator fields, and focus on the bosonic
SM multiplets acting as currents that probe the new sector.  The currents can
be introduced as local operators that couple to an unknown new-physics
spectrum in a manifestly $SU(2)_L\times U(1)_Y$ invariant way.  New dynamics
may involve weakly coupled (comparatively) light degrees of freedom, such as
extra Higgs singlets or doublets, it may probe a strongly coupled sector
which is resolved at high energy, or it may give rise to heavy
resonances, to name a few possibilities.  In any case, new physics that is
coupled to SM bosons, will manifest itself in anomalous scattering matrix
elements of those bosons, and should become accessible in high-energy VBS.  As
a common feature of this class of models, we expect the scattering matrix to
be self-contained and complete in terms of SM bosons and eventual new-physics
states, to a good approximation.\footnote{Heavy fermions ($t$, $b$, $\tau$)
  should 
  be taken into account in this context, but we do not study the corresponding
  final states in this paper.}

For a quantitative representation, we adopt the assumption of gauge invariance
and describe new physics as coupled to gauge-covariant monomials of SM fields.
For the building blocks, we introduce the Higgs multiplet in form of a
$2\times 2$ matrix,
\begin{align}
  \vH &=
  \frac 1 2
  \begin{pmatrix}
    v+ h -\ii w^3 & -\ii \sqrt{2} w^+ \\
		-\ii \sqrt{2} w^- & v + h + \ii w^3  \\
  \end{pmatrix}.
\end{align}
The components $h,w^\pm,w^3$ are the physical Higgs and unphysical
Goldstone scalars, respectively, and $v$ denotes the numerical Higgs
vev, $v=246\;\GeV$.  The matrix notation allows us to manifestly
represent the larger global symmetry on the Higgs field, $O(4)\sim
SU(2)_L\times SU(2)_R$ which after electroweak symmetry breaking
(EWSB) becomes the approximate custodial $SU(2)_c$ symmetry.
$SU(2)_L\times SU(2)_R$-symmetric monomials are invariant under bi-unitary
transformations 
of the form $\vH \to U_L \vH U_R^\dagger$.

The covariant derivative of the Higgs matrix is defined as
\begin{align}
  \label{eq:def-covariantDerivative}
  \vD_\mu \vH = \partial_\mu \vH - \ii g \vW_\mu \vH + \ii g^\prime \vH \vB_\mu \, ,
\end{align}
where 
\begin{align}
  \label{eq:def-gaugeFields}
  \vW_\mu \equiv \vW_\mu^a \frac{\tau^a}{2}, \qquad 
  \vB_\mu \equiv \vB_\mu^a \frac{\tau^3}{2}.
\end{align}
The transformation of $\vW_\mu$ is $\vW_\mu\to U_L^\dagger\vW_\mu U_L$, while
$\vB_\mu$ transforms covariantly only under a $U(1)_R$ subgroup of $SU(2)_R$.
The matrix-valued field strengths are given by
\begin{align}
  \label{eq:def-fieldStrength}
  \vW_{\mu\nu} = \partial_\mu \vW_\nu - \partial_\nu \vW_\nu - i g
  \left[ \vW_\mu, \vW_\nu \right], \qquad 
  \vB_{\mu\nu} = \partial_\mu \vB_\nu - \partial_\nu \vB_\nu \, .
\end{align}

From these fields, we can build local composite operators which act as
currents that probe the new, possibly non-local dynamics.  For instance, the
simplest Higgs-field currents are
\begin{align}
  J_H^{(2)} &= \tr{\vH^\dagger\vH},
  \label{eq:H0-current}
  &
  J_H^{(4)} &= \tr{(\vD_\mu\vH)^\dagger(\vD^\mu\vH)}
  &
  J_{H\mu\nu}^{(4)} &= \tr{(\vD_\mu\vH)^\dagger(\vD_\nu\vH)},
\end{align}
while gauge-field tensors can be combined as
\begin{align}
  J_W^{(4)} &=
  g^2\tr{\vW_{\mu\nu}\vW^{\mu\nu}},
  \label{eq:W0-current}
  &
  J_B^{(4)} &=
  g'{}^2\tr{\vB_{\mu\nu}\vB^{\mu\nu}},
  \\
  J_{W\mu\nu}^{(4)} &=
  g^2\tr{\vW_{\mu\rho}\vW^\rho_{\;\nu}},
  \label{eq:W2-current}
  &
  J_{B\mu\nu}^{(4)} &=
  g'{}^2\tr{\vB_{\mu\rho}\vB^\rho_{\;\nu}}.
\end{align}
These terms are electroweak singlets.  Non-singlet currents can
likewise be constructed.

We expect only weak constraints
from existing data, so new dynamics, whether parameterized by form factors,
spectral functions, or inelastic scattering into new particles, is rather
arbitrary.  For the purpose of this work, we focus on two extreme scenarios:
(i) a spectrum that interpolates the low-energy description with unitarity
saturation in the high-energy range, and (ii) a spectrum that consists of
separate narrow to medium-width resonances, which we may reduce to the
lowest-lying state for simplicity.  For reference, we also include (iii) the
unmodified SM where any new spectral functions are zero and all amplitudes
remain weakly interacting.  In terms of quasi-elastic $2\to 2$ scattering,
scenarios (i) and (iii) correspond to maximal and minimal event yields in the
asymptotic region, while (ii) exhibits unitarity saturation at finite energy.
Another extreme scenario, saturation by inelastic scattering into new final
states, asymptotically implies quasi-elastic event rates between (i) and (iii)
and should furthermore be accessible via direct observation of new particles.

As a first step, we may confine the analysis to pure Higgs- and
Goldstone-boson scattering.  This was done in our earlier
paper~\cite{Kilian:2014zja}.  Such a restriction implies further assumptions
on the underlying complete theory.  In this work, we remove this restriction.
We investigate the bosonic $2\to 2$ scattering matrix with Higgs,
longitudinal, and transversal vector bosons included.

Unless there are undetected light particles hiding in this scattering matrix,
it allows for a local operator-product expansion.  Contracting the singlet
currents listed above and ignoring terms which merely renormalize SM
parameters, the leading terms are dimension-six operators: $(J_H^{(2)})^3$,
$J_H^{(2)}J_H^{(4)}$, and $J_H^{(2)}J_W^{(4)}$.  Only the latter term is
easily accessible at the LHC,\footnote{There are two directions in the
  dimension-six operator space which in our context correspond to
  $(J_H^{(2)})^3$ and $J_H^{(2)}J_H^{(4)}$.  An unambiguous determination of the
  coefficients requires measurements of Higgs-pair production and the Higgs
  total width, respectively.} 
so a phenomenological parameterization should consider the next order of the
expansion.  These are dimension-eight local interactions.

Including all singlet and non-singlet operator products, and omitting CP-odd
interactions, we can identify three distinct categories of dimension-eight
bosonic operators in the low-energy expansion that we list below.

There are two terms which couple only Higgs-field currents,
\begin{subequations}
\label{e:longlang}
\begin{align}
\mathcal{L}_{S,0}&=F_{S,0}\text{tr}\left[(\textbf{D}_\mu
\textbf{H})^\dagger (\textbf{D}_\nu \textbf{H})\right]
\text{tr}\left[(\textbf{D}^\mu \textbf{H})^\dagger
(\textbf{D}^\nu \textbf{H})\right], \\
\mathcal{L}_{S,1}&=F_{S,1}\text{tr}\left[(\textbf{D}_\mu
\textbf{H})^\dagger (\textbf{D}^\mu \textbf{H})\right]
\text{tr}\left[(\textbf{D}_\nu \textbf{H})^\dagger
(\textbf{D}^\nu \textbf{H})\right] \; ;
\end{align}
\end{subequations}
seven terms which couple Higgs- and gauge field currents,
\begin{subequations}
\label{e:mixlang}
\begin{align}
\mathcal{L}_{M,0}&=-g^2 F_{M_0}\text{tr}\left[(\textbf{D}_\mu
\textbf{H})^\dagger (\textbf{D}^\mu \textbf{H})\right]
\text{tr}\left[\textbf{W}_{\nu \rho} \textbf{W}^{\nu \rho}\right]
, \\
\mathcal{L}_{M,1}&=-g^2 F_{M_1}\text{tr}\left[(\textbf{D}_\mu
\textbf{H})^\dagger (\textbf{D}^\rho \textbf{H})\right]
\text{tr}\left[\textbf{W}_{\nu \rho} \textbf{W}^{\nu \mu}\right],
\\
\mathcal{L}_{M,2}&=-g^{\prime 2} F_{M_2}\text{tr}\left[(\textbf{D}_\mu
 \textbf{H})^\dagger (\textbf{D}^\mu \textbf{H})\right] \text{tr}
 \left[\textbf{B}_{\nu \rho} \textbf{B}^{\nu \rho}\right], \\
\mathcal{L}_{M,3}&=-g^{\prime 2} F_{M_3}\text{tr}\left[(\textbf{D}_\mu
\textbf{H})^\dagger (\textbf{D}^\rho \textbf{H})\right] \text{tr}
\left[\textbf{B}_{\nu \rho} \textbf{B}^{\nu \mu}\right], \\
\mathcal{L}_{M,4}&=-g g^\prime  F_{M_4}\text{tr}\left[(\textbf{D}_\mu
\textbf{H})^\dagger \textbf{W}_{\nu \rho} (\textbf{D}^\mu \textbf{H})
\textbf{B}^{\nu \rho}\right], \\
\mathcal{L}_{M,5}&=-g g^\prime  F_{M_5}\text{tr}\left[(\textbf{D}_\mu
\textbf{H})^\dagger \textbf{W}_{\nu \rho} (\textbf{D}^\rho \textbf{H})
\textbf{B}^{\nu \mu}\right], \\
\mathcal{L}_{M,7}&=-g^2 F_{M_7}\text{tr}\left[(\textbf{D}_\mu
\textbf{H})^\dagger \textbf{W}_{\nu \rho} \textbf{W}^{\nu \mu}
(\textbf{D}^\rho \textbf{H}) \right] \; ;
\end{align}
\end{subequations}
and eight terms which couple gauge-field currents to themselves:
\begin{subequations}
\label{e:translang}
\begin{align}
\mathcal{L}_{T_0}&=g^4 F_{T_0}\text{tr}\left[\textbf{W}_{\mu \nu}
\textbf{W}^{\mu \nu}\right]
\text{tr}\left[\textbf{W}_{\alpha \beta} \textbf{W}^{\alpha \beta}
\right], \\
\mathcal{L}_{T_1}&=g^4 F_{T_1}\text{tr}\left[\textbf{W}_{\alpha \nu}
\textbf{W}^{\mu \beta}\right]
\text{tr}\left[\textbf{W}_{\mu \beta} \textbf{W}^{\alpha \nu}\right],
\\
\mathcal{L}_{T_2}&=g^4 F_{T_2}\text{tr}\left[\textbf{W}_{\alpha \mu}
\textbf{W}^{\mu \beta}\right]
\text{tr}\left[\textbf{W}_{\beta \nu} \textbf{W}^{\nu \alpha}\right],
\\
\mathcal{L}_{T_5}&=g^2 g^{\prime 2} F_{T_5}\text{tr}
\left[\textbf{W}_{\mu \nu} \textbf{W}^{\mu \nu}\right]
\text{tr}\left[\textbf{B}_{\alpha \beta} \textbf{B}^{\alpha \beta}\right],
\\
\mathcal{L}_{T_6}&=g^2 g^{\prime 2} F_{T_6}\text{tr}
\left[\textbf{W}_{\alpha \nu} \textbf{W}^{\mu \beta}\right]
\text{tr}\left[\textbf{B}_{\mu \beta} \textbf{B}^{\alpha \nu}\right], \\
\mathcal{L}_{T_7}&=g^2 g^{\prime 2} F_{T_7}\text{tr}
\left[\textbf{W}_{\alpha \mu} \textbf{W}^{\mu \beta}\right]
\text{tr}\left[\textbf{B}_{\beta \nu} \textbf{B}^{\nu \alpha}\right], \\
\mathcal{L}_{T_8}&=g^{\prime 4} F_{T_8}\text{tr}\left[\textbf{B}_{\mu \nu}
\textbf{B}^{\mu \nu}\right]
\text{tr}\left[\textbf{B}_{\alpha \beta} \textbf{B}^{\alpha \beta}\right],
\\
\mathcal{L}_{T_9}&=g^{\prime 4} F_{T_9}\text{tr}\left[\textbf{B}_{\alpha \mu}
 \textbf{B}^{\mu \beta}\right]
\text{tr}\left[\textbf{B}_{\beta \nu} \textbf{B}^{\nu \alpha}\right].
\end{align}
\end{subequations}

Note that the enumeration of operators is not
consecutive. We have adopted the naming convention from the
literature~\cite{Eboli:2006wa,Degrande:2013kka,Baak:2013fwa} but eliminated
redundant interactions to arrive at a linearly independent set.

We emphasize that this list only describes the model-independent low-energy
limit of the true amplitude.  The actual measurement of VBS processes is not
restricted to the low-energy range and thus cannot be accurately accounted for
by the low-energy limit only.  The true quasi-elastic scattering amplitudes
will resolve the local operators into non-local interactions and thus keep the
result in accordance with the applicable unitarity relations at all energies.
This is the rationale for introducing simplified models such as (i) and (ii)
above.

For the purpose of this study, we adopt a simplification that applies to all
considered models: we impose 
global custodial symmetry on the beyond the SM (BSM) interactions and thus
omit all terms that involve the hypercharge boson.  In the local operator
basis, we are left with two parameters $F_{S_{0/1}}$, three parameters
$F_{M_{0/1/7}}$, and three parameters $F_{T_{0/1/2}}$.  This choice implies
that $W$ and $Z$ amplitudes are mutually related, and that photon interactions
do not carry independent information.

\section{Properties of the Scattering Matrix}
\label{sec:diagonal}

We apply the phenomenological description of the preceding section to the set
of VBS scattering amplitudes, which we then embed in complete LHC processes.
The basic processes are all of the $2\to 2$ quasi-elastic scattering type.  In
this situation, standard scattering theory applies.  We may evaluate
partial-wave amplitudes and thus convert the scattering amplitudes to a
finite-dimensional matrix.  This allows us to diagonalize the scattering
matrix and find a unitary projection of each eigenamplitude individually, if
the calculated model amplitude does not respect partial-wave unitarity.

This simplification is based on approximations.  We ignore external and
internal photons.  This implies the custodial-$SU(2)$ limit, as already
discussed in the previous section. It also implies that we ignore the Coulomb
pole in charged-$W$ scattering.  The omission is justified since the forward
region is cut out in an experimental analysis, while the Coulomb singularity
is not reached for the complete process, due to the spacelike nature of the
incoming virtual vector bosons.  We also treat the external particles as
on-shell, while in the real process at the LHC, the initial vector bosons are
actually space-like.  In effect, these omissions amount to subleading
corrections of relative order $m_W^2/\hat s$ and ${q_i^2}/{\hat{s}}$ for the
$2\to 2$ quasi-elastic scattering processes, where $q_i$ is the space-like
momentum of an incoming vector boson.  We note that for VBS kinematics, values
$|q_i^2|\sim m_W^2$ dominate the cross section, but there is a phase-space
region where terms proportional to ${q^2}/{\hat{s}}$ become leading.  In the
current paper, we focus on observables inclusive in $q^2$ where these terms
are mostly subleading.  We refer to Ref.~\cite{Perez:2018} for a more
exhaustive discussion.

In fact, the symmetry structure of our simplified models allows for a
choice of basis that renders the scattering matrix diagonal at all
energies, up to subleading corrections.  Asymptotically, the
longitudinal vector boson modes combine with the Higgs mode, while the
transverse modes decouple.  The external states combine to multiplets of the
custodial $SU(2)$ symmetry.  
This property is well known for the SM.  If we assume custodial symmetry also
for the new interactions, we can use it for expressing all quasi-elastic
scattering amplitudes of Higgs and longitudinal vector boson modes in terms of
a single scalar 
master amplitude, which can be used to find partial-wave eigenamplitudes and
their unitary
projection~\cite{Alboteanu:2008my,Sekulla:2015enc,Kilian:2014zja}.  Here, we
apply the same principle to transverse and mixed scattering amplitudes.

The key observation is that the contact interactions of the
SMEFT~(\ref{e:longlang}, \ref{e:mixlang}, \ref{e:translang}), although they do
not provide a satisfactory phenomenological description, already
encode the most general dependence of the scattering matrix on
external quantum numbers, if we
restrict the analysis to quasi-elastic $2\to 2$ scattering.  To
describe an arbitrary new-physics spectrum, not just the low-energy
limit, we merely have to promote the coefficients $F_i$ to scalar
form-factors which can depend on $s,t,u$.  Turning this around, we can
formally diagonalize the scattering matrix in terms of those
coefficients.  Unlike the scattering matrix for longitudinal modes
only, this procedure involves the helicities of the external vector
bosons $\lambda_i$.  Since the procedure is required only for the high-energy
range, we neglect the masses of $W$, $Z$, and $H$ where applicable.

For the calculation below, we can thus treat the non-SM part of the amplitudes
as if they were given by the local dimension-eight operator approximation,
keeping in mind that the method works as well for non-constant coefficients.
The unitary projection that we obtain assumes the same form, with specific
functions for the coefficients, and applying the same projection a second time
will not change the asymptotic form of the result anymore.

For the transverse interactions with structures
$\mathcal{L}_{T_{0/1/2}}$ and for the mixed interactions 
with the operators $\mathcal{L}_{M_{0/1/7}}$ we define the master amplitude
\begin{align}
\label{e:trans_master}
A(s,t,u;\lambda_1,\lambda_2,\lambda_3,\lambda_4)=&\mathcal{A}
(W^+_{\lambda_1} W^-_{\lambda_2} \to Z_{\lambda_3} Z_{\lambda_4})
\notag \\
=&-2g^4 (F_{T_0}+\frac{1}{4}F_{T_2}) \delta_{\lambda_1,\lambda_2}
\delta_{\lambda_3,\lambda_4}s^2 \notag \\
&- g^4 (F_{T_1}+\frac{1}{2}F_{T_2})\left( \delta_{\lambda_1,-\lambda_3}
\delta_{\lambda_2,-\lambda_4}t^2+ \delta_{\lambda_1,-\lambda_4}
\delta_{\lambda_2,-\lambda_3}u^2 \right) \notag\\
&+\frac{1}{2}g^4 F_{T_2} \delta_{\lambda_1,\lambda_2} \delta_{\lambda_3,
\lambda_4} \delta_{\lambda_1,-\lambda_3}(t^2+u^2) \notag \\
&+ \frac{1}{16} g^2 (8 F_{M_0} - 2 F_{M_1} + F_{M_7}) \left(
\delta_{\lambda_1,\lambda_2}
\delta_{\lambda_3,0} \delta_{\lambda_4,0} - \delta_{\lambda_3,\lambda_4}
16\delta_{\lambda_1,0} \delta_{\lambda_2,0} \right) s^2 \notag \\
&+ \frac{1}{16} g^2 (2 F_{M_1} + F_{M_7})
\left( \delta_{\lambda_1,-\lambda_2}
\delta_{\lambda_3,0} \delta_{\lambda_4,0} -
\delta_{\lambda_3,-\lambda_4} \delta_{\lambda_1,0} \delta_{\lambda_2,0}
\right) \left( s^2 - t^2 - u^2 \right) \notag \\
&+ \frac{1}{16} F_{M_7} \biggl[ \left( \delta_{\lambda_1,-\lambda_3}
\delta_{\lambda_2,0} \delta_{\lambda_4,0} -\delta_{\lambda_2,-\lambda_4}
\delta_{\lambda_1,0} \delta_{\lambda_3,0} \right) \left (s^2 - u^2 \right)
\notag  \\
& \qquad\qquad+  \left( \delta_{\lambda_2,-\lambda_3}
\delta_{\lambda_1,0} \delta_{\lambda_4,0} -\delta_{\lambda_1,-\lambda_4}
\delta_{\lambda_2,0} \delta_{\lambda_3,0} \right) \left (s^2 - t^2 \right)
\biggr].
\end{align}

The decomposition of the scattering amplitudes into isospin
eigenamplitudes is identical for mixed and transverse operators and
given by
\begin{subequations}
\label{e:ampwithiso}
\begin{align}
\mathcal{A}(W^+_{\lambda_1} W^+_{\lambda_2} \to W^+_{\lambda_3}
W^+_{\lambda_4})&=\phantom{\frac{1}{3}}\mathcal{A}_2 (s,t,u;\boldsymbol{\lambda})\\
\mathcal{A}(W^+_{\lambda_1} W^-_{\lambda_2} \to W^+_{\lambda_3}
W^-_{\lambda_4})&=\frac{1}{3}\mathcal{A}_0 (s,t,u;\boldsymbol{\lambda})+
\frac{1}{2}\mathcal{A}_1 (s,t,u;\boldsymbol{\lambda})+\frac{1}{6}
\mathcal{A}_2 (s,t,u;\boldsymbol{\lambda})\\
\mathcal{A}(W^+_{\lambda_1} W^-_{\lambda_2} \to Z_{\lambda_3}
Z_{\lambda_4})&=\frac{1}{3}\mathcal{A}_0 (s,t,u;\boldsymbol{\lambda})-
\frac{1}{3}\mathcal{A}_2 (s,t,u;\boldsymbol{\lambda})\\
\mathcal{A}(W^+_{\lambda_1} Z_{\lambda_2} \to W^+_{\lambda_3}
Z_{\lambda_4})&=\frac{1}{2}\mathcal{A}_1 (s,t,u;\boldsymbol{\lambda})+
\frac{1}{2}\mathcal{A}_2 (s,t,u;\boldsymbol{\lambda})\\
\mathcal{A}(Z_{\lambda_1} Z_{\lambda_2} \to Z_{\lambda_3}
Z_{\lambda_4})&=\frac{1}{3}\mathcal{A}_0 (s,t,u;\boldsymbol{\lambda})+
\frac{2}{3}\mathcal{A}_2 (s,t,u;\boldsymbol{\lambda}) \; .
\end{align}
\end{subequations}
Here, $\boldsymbol \lambda =(\lambda_1,\lambda_2,\lambda_3,\lambda_4)$
is a multi-index for the four different helicities of the weak vector
bosons. Using this, the isospin eigenamplitudes are given by
\begin{subequations}
\label{e:isospinamplitudes}
\begin{align}
\mathcal{A}_0(s,t,u;\boldsymbol{\lambda}) = & \quad\;
3A(s,t,u;\lambda_1,\lambda_2,\lambda_3,\lambda_4) \notag\\
&+A(t,s,u;-\lambda_4,\lambda_2,\lambda_3,-\lambda_1) \notag \\*
&+A(u,t,s;\lambda_1,-\lambda_4,\lambda_3,-\lambda_2)\\
\mathcal{A}_1(s,t,u;\boldsymbol{\lambda}) = & \quad\;
A(t,s,u;-\lambda_4,\lambda_2,\lambda_3,-\lambda_1) \notag \\*
&-A(u,t,s;\lambda_1,-\lambda_4,\lambda_3,-\lambda_2)\\
\mathcal{A}_2(s,t,u;\boldsymbol{\lambda}) = & \quad\;
A(t,s,u;-\lambda_4,\lambda_2,\lambda_3,-\lambda_1) \notag \\*
&+A(u,t,s;\lambda_1,-\lambda_4,\lambda_3,-\lambda_2).
\end{align}
\end{subequations}

The next step is the decomposition into isospin-spin eigenamplitudes
which is done by the expansion of the isospin eigenamplitudes
(\ref{e:ampwithiso}) into the Wigner D-functions~\cite{Wigner:1931}
$d^J_{\lambda,\lambda^\prime}(\theta)$ with $\lambda=\lambda_1-\lambda_2$
and $\lambda^\prime=\lambda_3-\lambda_4$.
\begin{align}
\label{e:isospinspincalc}
\mathcal{A}_{IJ}(s;\boldsymbol{\lambda})=\int_{-s}^0\frac{dt}{s}
A_I(s,t,u;\boldsymbol{\lambda}) \cdot d^J_{\lambda,\lambda^\prime}\left[\arccos
\left(1+2\frac{t}{s}\right)\right]
\end{align}
Tables~\ref{t:isospin-spin-coeff_m} (transverse operators)
and~\ref{t:isospin-spin-coeff_t} (mixed operators) list the complete set of
master amplitudes with their dependence on the operator coefficients, i.e.,
the asymptotically leading behavior.  These helicity-dependent eigenamplitudes
can in principle be used for determining unitary projections as form factors
that multiply modified Feynman rules for the boson fields.

For the purpose of constructing a minimal unitary projection, it is sufficient
to determine a set of master amplitudes which capture the leading term
proportional to $s^2$ for each spin-isospin channel, uniformly for all
individual helicity combinations.  The implied over-compensation of some
helicity channels that are subleading at high energy, is within the scheme
dependence that is inherent in the unitary projection.  We find the following
simplified, helicity-independent expressions:
\begin{subequations}
\label{e:isospinspin_trans}
\begin{align}
\mathcal{A}_{00}(s) =& - \frac{3}{2} g^4 \left[ 4 F_{T_0} - 2 F_{T_1} +
F_{T_2}\right] s^2  \nonumber\\
& + \frac{3}{16} g^2 \left[ 8 F_{M_0} + 2 F_{M_1} +
F_{M_7} \right ]  s^2
\\
\mathcal{A}_{01}(s) =& - \frac{1}{32} g^2 \left[ 4 F_{M_0} + F_{M_1} -
3 F_{M_7} \right] s^2\\
\mathcal{A}_{02}(s) =& - \frac{1}{10} g^4 \left[4 F_{T_0} - 2 F_{T_1} +
F_{T_2}\right] s^2 \nonumber\\
&+ \frac{1}{160} g^2 \left[ 4 F_{M_0} + F_{M_1} +
F_{M_7} \right] s^2 \\ 
\mathcal{A}_{10}(s) =& 0 \\
\mathcal{A}_{11}(s) =&  - \frac{1}{6} g^4 F_{T_2} s^2 \nonumber\\
& - \frac{1}{32} g^2 \left[ 4 F_{M_0} + F_{M_1} -
3 F_{M_7}\right] s^2 \\ 
\mathcal{A}_{12}(s) =& \frac{1}{5} g^4 \left[- 2 F_{T_0} + F_{T_1}\right] s^2 \nonumber\\
& + \frac{1}{160} g^2 \left[ 4 F_{M_0} + F_{M_1} +
F_{M_7} \right] s^2 \\
\mathcal{A}_{20}(s) =& 0\\
\mathcal{A}_{21}(s) =& - \frac{1}{32} g^2 \left[ 4 F_{M_0} + F_{M_1} -
3 F_{M_7} \right] s^2\\
\mathcal{A}_{22}(s) =& - \frac{1}{10} g^4 \left[4 F_{T_0} - 2 F_{T_1} +
F_{T_2}\right] s^2 \nonumber\\
& + \frac{1}{160} g^2 \left[ 4 F_{M_0} + F_{M_1} +
F_{M_7} \right] s^2  \, .
\end{align}
\end{subequations}
In fact, comparing polarized $2\to 2$ on-shell processes we have verified that
the numerical discrepancy between a helicity-dependent treatment and the
simplified version is less than a percent for the parameter ranges considered,
negligible in comparison to the scheme dependence of the unitary projection
itself.

We now turn to the unitary projection of the scattering amplitudes, applicable
to the high-energy range where the leading behavior in the presence of nonzero
operator coefficients is given by~(\ref{e:isospinspin_trans}).  We follow the
T-matrix projection scheme introduced in Ref.~\cite{Kilian:2014zja} and apply
it to the simplified helicity-independent eigenamplitudes, for the case where
only one type of new interactions (mixed or transversal) is active at a time.  The simplifications
combined allow us to evaluate
the projection and thus the compensating terms in closed form.  (If
coefficients are non-zero for both classes simultaneously or a detailed
separation of 
helicities is intended, we have to resort to numerical evaluation of
the T-matrix projection.  This is beyond the scope of the present work.)

The unitary projection of a spin-isospin eigenamplitude ${\mathcal{A}}_{IJ}$
is given by the expression
\begin{align}
\label{e:isospinspin_unit}
\hat{\mathcal{A}}_{IJ} = \frac{1}{ \mathrm{Re} {\frac{1}{\mathcal{A}_{IJ}}}- \frac{\ii}{32 \pi}}.
\end{align}
This projection may be recast as an $s$-dependent correction
counterterm for each eigenamplitude,
\begin{align}
\label{e:unit_corr}
\Delta \mathcal{A}_{IJ} = \hat{\mathcal{A}}_{IJ} - \mathcal{A}_{IJ}.
\end{align}
The limit $A_{IJ}\to\infty$ lets us recover the universal
unitarity bound for each eigenamplitude,
\begin{align}
  |\hat{\mathcal{A}}_{IJ}| \leq 32 \pi \, .
\end{align}
In particular, the truncated SMEFT expansion, i.e. constant coefficients in
the eigenamplitude above, yields $\lim_{s\to\infty}\mathcal{A}_{IJ}\to\infty$
for all partial waves with nonzero coefficients.  The T-matrix projection then
asymptotically saturates unitarity.  A model with a pole in the amplitude at
some value $s=M^2$, projected according to this prescription, saturates the
unitarity limit at this point and follows a Breit-Wigner shape for the energy
dependence in the vicinity of the pole.  The actual pole of the amplitude gets
shifted away from the real axis.

\section{(Strongly coupled) Continuum Model}
\label{sec:contmodel}

As mentioned in Sec.~\ref{sec:currents}, the simplified models that we
actually consider are (i) continuum models which smoothly interpolate
between high-energy unitarity saturation and the low-energy SMEFT, and
(ii) resonance models where distinct features arise in the spectrum.
Including also the unmodified SM in the discussion, these models cover the
whole range 
of possible interaction strengths that future VBS measurements may
observe, and thus yield a fairly robust projection for the sensitivity
of a collider experiment.  With the exception of the unmodified SM, neither of
these models is UV complete, and the actual results should behave
differently in the asymptotic regime.  For instance, new inelastic
channels may appear as final states.  However, as long as the initial
assumptions about unitarity, gauge invariance and minimal flavor
violation hold true, we should not expect event rates in this sector
which exceed the strongly-interacting continuum scenario that we
consider here.

For our numerical studies of the continuum scenario, we have adopted
the amplitudes with the local operators of Sec.~\ref{sec:currents}
added to the SM Feynman rules and converted this to a unitary model
according to~\eqref{e:isospinspin_unit} after diagonalization.  This
has been re-expressed in terms of form-factor modified Feynman rules along
the lines of Refs.~\cite{Alboteanu:2008my,Kilian:2014zja} and
implemented in the Monte-Carlo event generator
\whizard~\cite{Kilian:2007gr}\footnote{\whizard\ is a multi-purpose
  event generator which ships with its a tree-level
  matrix element generator~\cite{Moretti:2001zz,Nejad:2014sqa}. It uses the
  color-flow formalism for QCD~\cite{Kilian:2012pz}, and comes with
  its own parton shower 
  implementations~\cite{Kilian:2011ka}. While it allows to simulate
  almost arbitrary BSM interactions (cf. e.g. the SUSY
  implementation~\cite{Ohl:2002jp}) via its \texttt{FeynRules}
  interface~\cite{Christensen:2010wz}, \whizard\ has been 
  also successfully extended towards next-to-leading order and matched
  to gluon and photon resummation~\cite{WHIZARD_NLO}}.  
Using Feynman rules and a
straightforward on-shell projection of the boson momenta, the
interactions can be evaluated off-shell in the context of an automatic
amplitude evaluation, and thus enter the standard \whizard\ framework that
ultimately yields simulated event samples.  There are some subtleties
hidden in the on-shell projection; this is discussed in detail, along
with some refinement of the method, in Ref.~\cite{Perez:2018}. 

\begin{figure}[!h]
   \begin{subfigure}[t]{0.5\textwidth}
      \includegraphics[width=\textwidth]{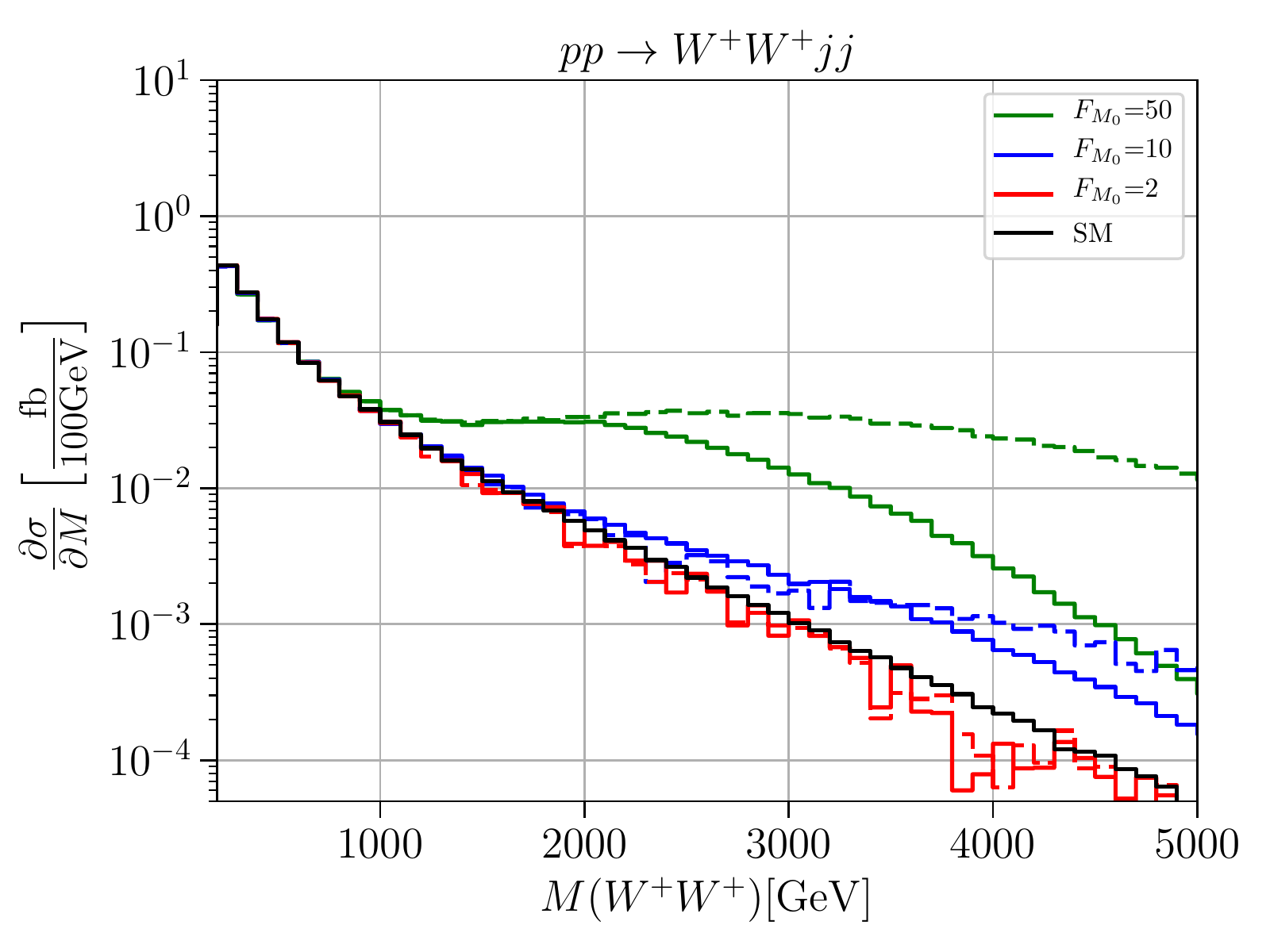}
   \end{subfigure}\hfill%
   \begin{subfigure}[t]{0.5\textwidth}
      \includegraphics[width=\textwidth]{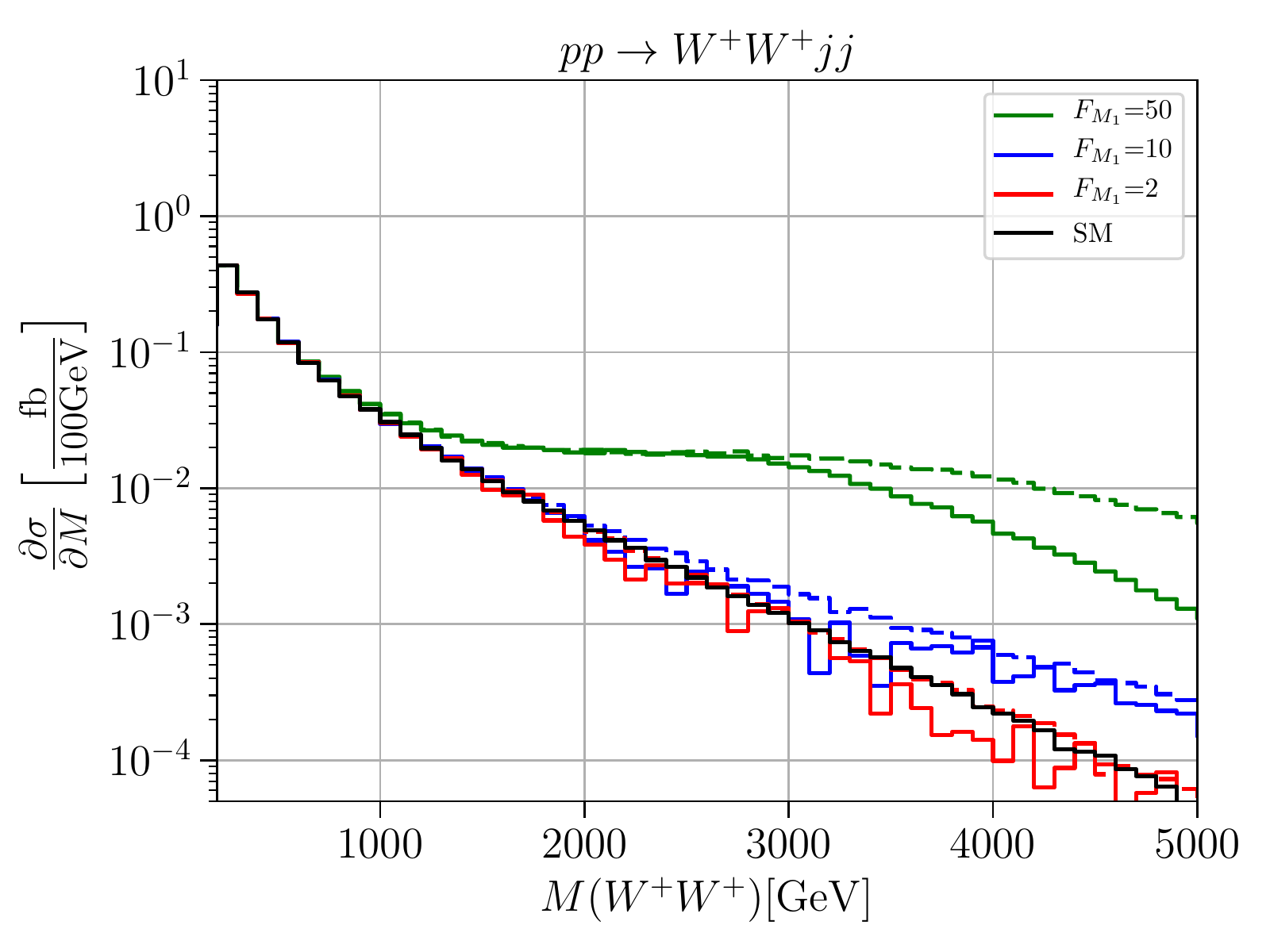}
   \end{subfigure}\\[5pt]%
   \centering
      \begin{subfigure}[t]{0.5\textwidth}
      \includegraphics[width=\textwidth]{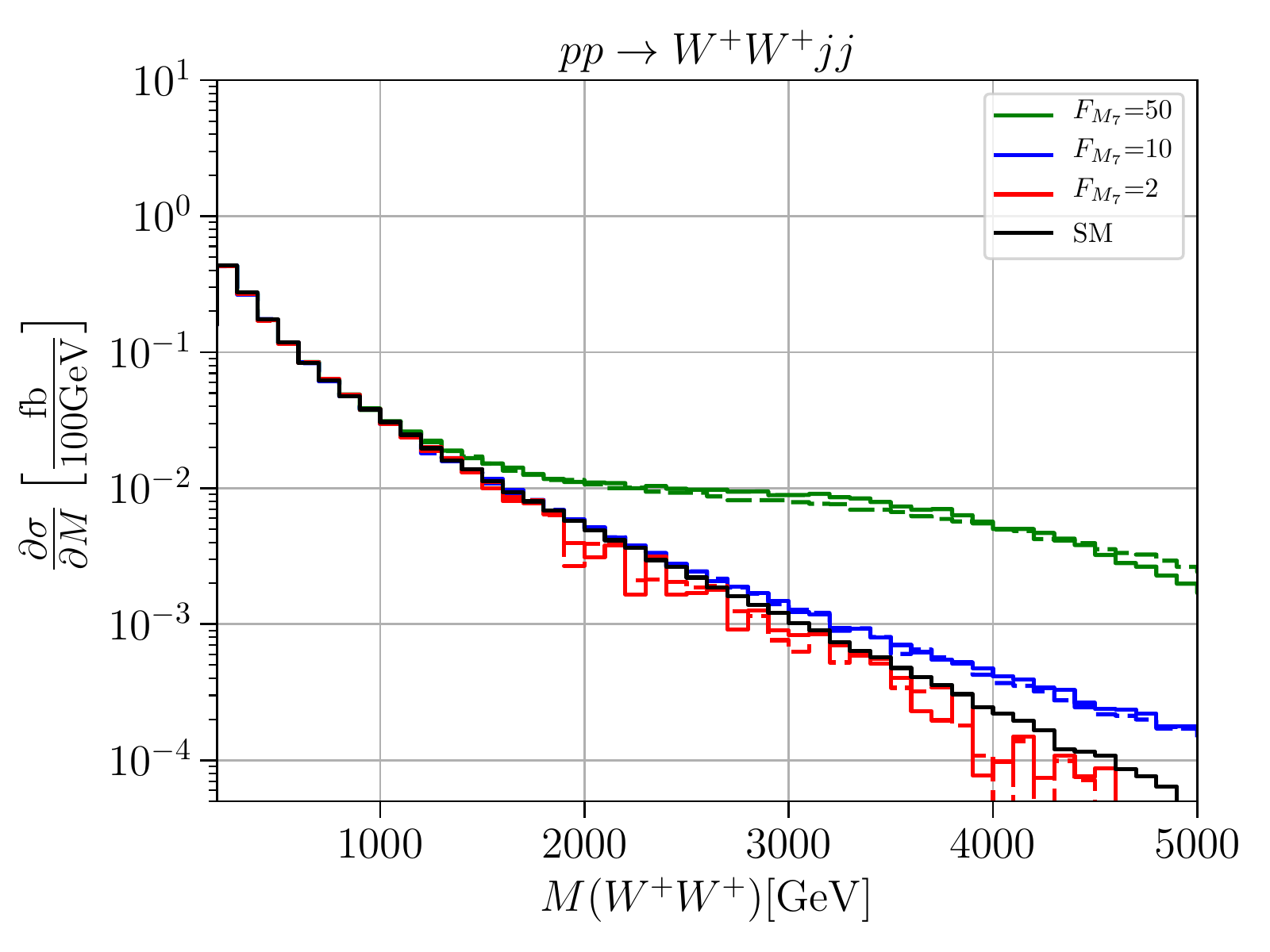}
   \end{subfigure}\\[5pt]%
   \caption{Cross sections differential in the diboson invariant mass
     for the process $pp \to W^+W^+jj$. The solid black line shows the
     Standard Model differential cross section, the green, blue and
     red lines the cross sections with anomalous couplings $F_{M_i}=50\,\
     \TeV^{-4}, F_{M_i}=10\,\TeV^{-4}$ and $F_{M_i}=2\,\TeV^{-4} $ for
     i = 0 (upper left panel), i = 1 (upper right panel), and i = 7
     (lower panel), respectively.
     Solid: unitarized; dashed: naive result.
     Cuts: $M_{jj}>500\,\ \GeV$, $\Delta \eta_{jj}>2.4$,
     $|\eta_j|<4.5$, $p_T^j>20\,\ \GeV$.}
   \label{i:FM_unit}
\end{figure}

\begin{figure}[!h]
  \begin{subfigure}[t]{0.5\textwidth}
     \includegraphics[width=\textwidth]{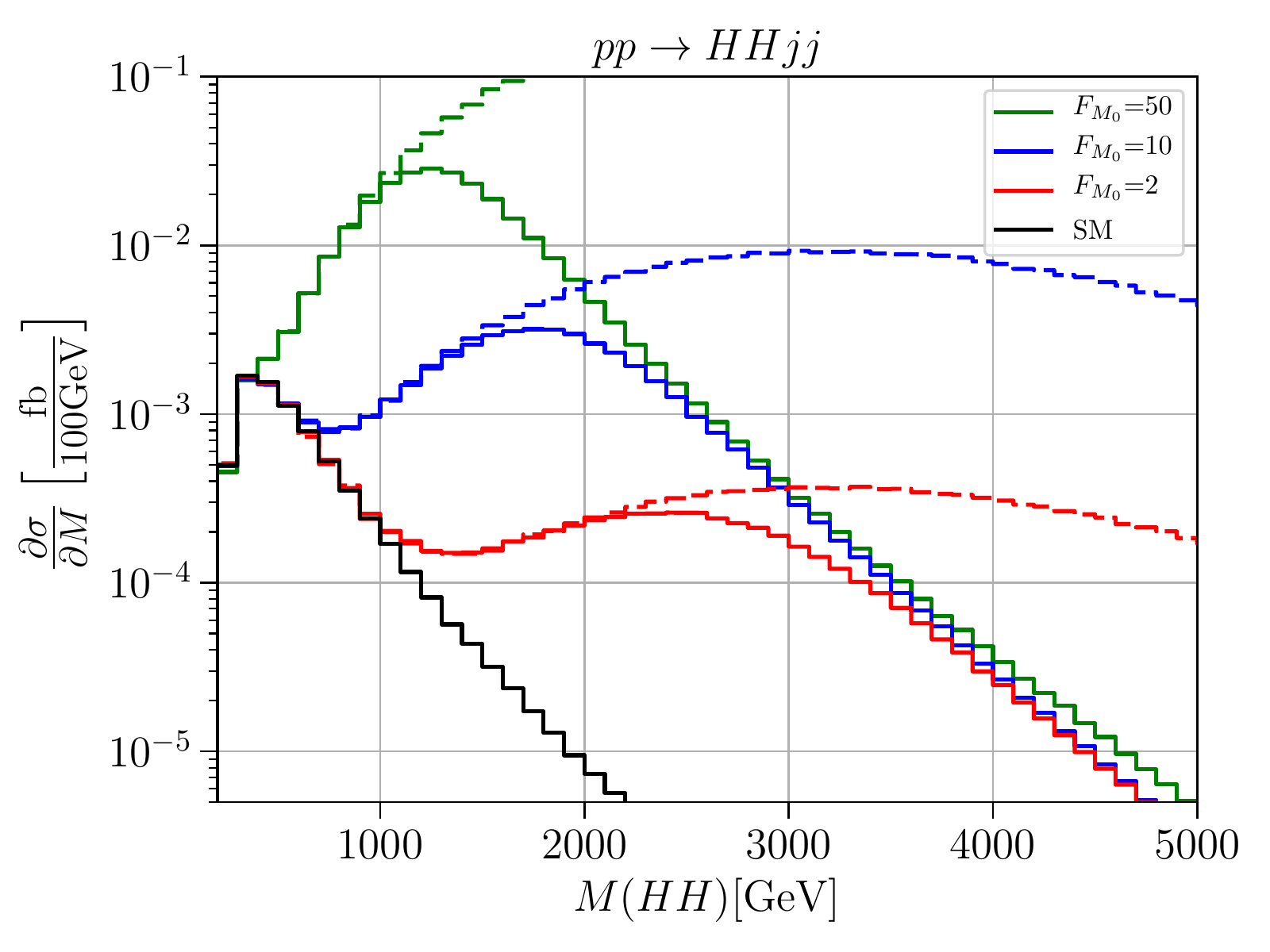}
  \end{subfigure}\hfill%
  \begin{subfigure}[t]{0.5\textwidth}
     \includegraphics[width=\textwidth]{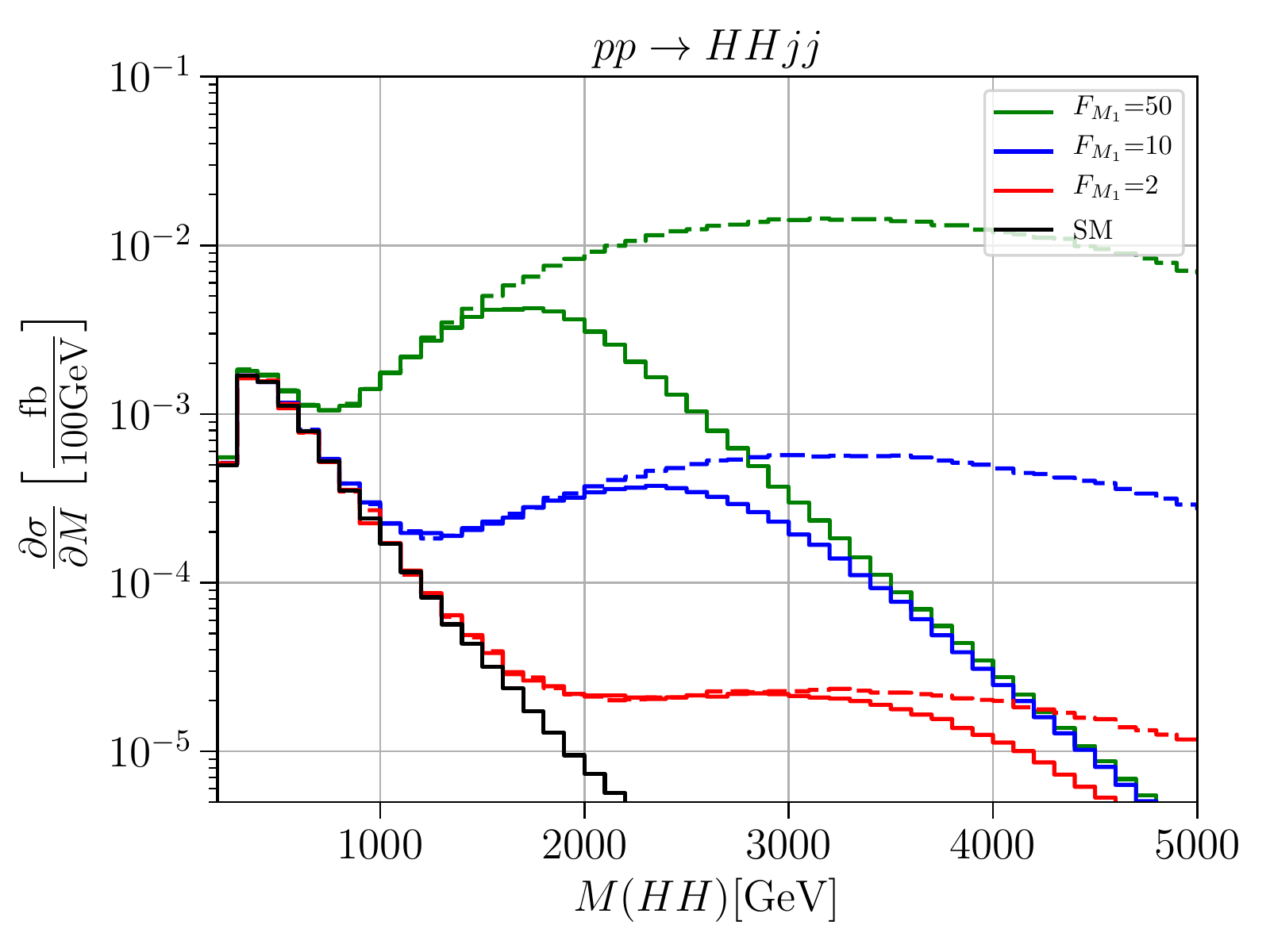}
  \end{subfigure}\\[5pt]%
  \centering
     \begin{subfigure}[t]{0.5\textwidth}
     \includegraphics[width=\textwidth]{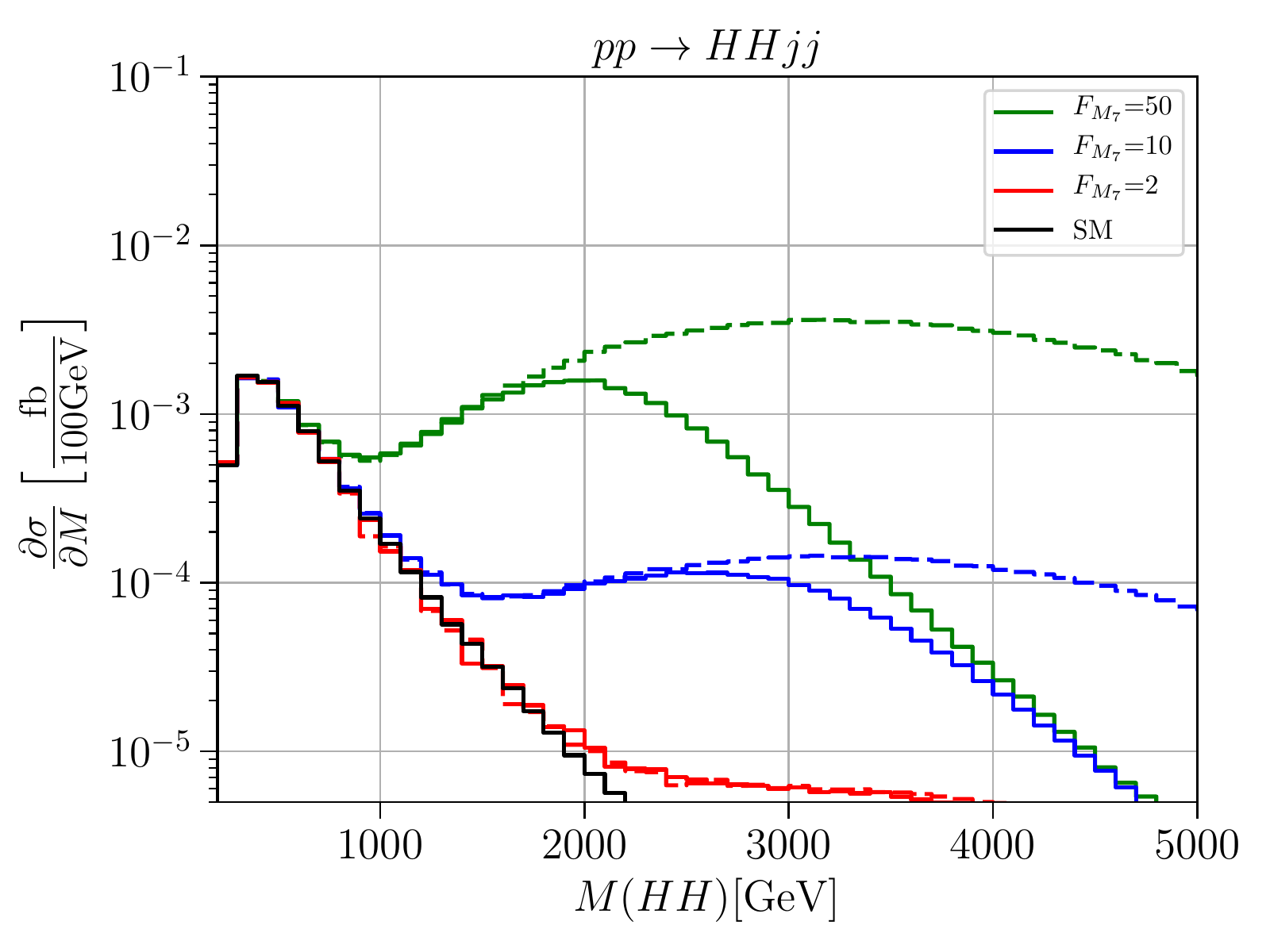}
  \end{subfigure}\\[5pt]%
  \caption{Cross sections differential in the diboson invariant mass
    for the process $pp \to HHjj$. The solid black line shows the
    Standard Model differential cross section, the green, blue and red lines the
    cross sections with anomalous couplings $F_{M_i}=50\,\
    \TeV^{-4}, F_{M_i}=10\,\TeV^{-4}$ and $F_{M_i}=2\,\TeV^{-4} $ for
    i = 0 (upper left panel), i = 1 (upper right panel), and i = 7
    (lower panel), respectively.
    Solid: unitarized; dashed: naive result.
    Cuts: $M_{jj}>500\,\ \GeV$, $\Delta \eta_{jj}>2.4$,
    $|\eta_j|<4.5$, $p_T^j>20\,\ \GeV$.}
  \label{i:FM_unit-HH}
\end{figure}

The processes $pp\to jjW^+W^+$ and $pp\to jjHH$ have received special
attention.  The former process exhibits a characteristic signature of
like-sign dileptons and has the largest signal-to-branching ratio of
all VBS processes at the LHC, while the latter is difficult to isolate
but carries a dependence on the triple-Higgs coupling which is among
the most elusive SM parameters.  In fact, an anomalous triple-Higgs
coupling can be attributed to a gauge-invariant dimension-six
operator, while in this work we are considering dimension-eight
contributions.  Clearly, an unambiguous determination of a
dimension-six parameter in a systematic low-energy expansion is only
possible if the next higher order is under control. 

In Fig.~\ref{i:FM_unit} and in Fig.~\ref{i:FM_unit-HH}
we show results for the process $pp \to W^+W^+
jj$ and for $pp \to HHjj$  within the continuum simplified
model with nonzero coefficients for the longitudinal-transverse mixed
operators with parameters $F_{M_{0/1/7}}$, respectively. 
We choose three distinct values, $F=2,10,50\;\TeV^{-4}$, with one
nonzero coefficient at a time.  The solid lines show the distribution
in the invariant mass of the $W^+W^+/HH$ pair, which coincides with the
effective energy $\sqrt{\hat s}$ for the basic VBS process.  We note
that in the presence of background and finite jet-energy resolution,
this distribution is not actually measurable for $W^+W^+$.  Rather,
$W$-boson decay leptons are detected.  However, the plots describe
most clearly the expected physics, which will only be diluted in
actual observables.

For reference, we also display the results which would be obtained if the
naive dimension-eight SMEFT amplitude, without T-matrix correction,
were used for the calculation (dashed).  Clearly, such a calculation
overestimates the achievable event yield by a huge amount and suggests
a sensitivity to the model parameters which is unphysical.

We observe that regardless of parameters, the solid curves approach an
asymptotic differential cross section which for the $W^+W^+$ final state is
enhanced by about an order of magnitude over the SM prediction. In the case of
the $HH$ final state, the enhancement amounts to more than two orders of
magnitude.  These asymptotic limits correspond to a maximally strong
interaction, saturation of the unitarity limit within the quasi-elastic
channel.  The residual parameter dependence is confined to a certain
transition region.  Beyond this region, from the saturated quasi-elastic
amplitudes we can read off the maximally allowed event number for the given
spin-isospin channel.



\begin{figure}[!h]
   \begin{subfigure}[t]{0.5\textwidth}
      \includegraphics[width=\textwidth]{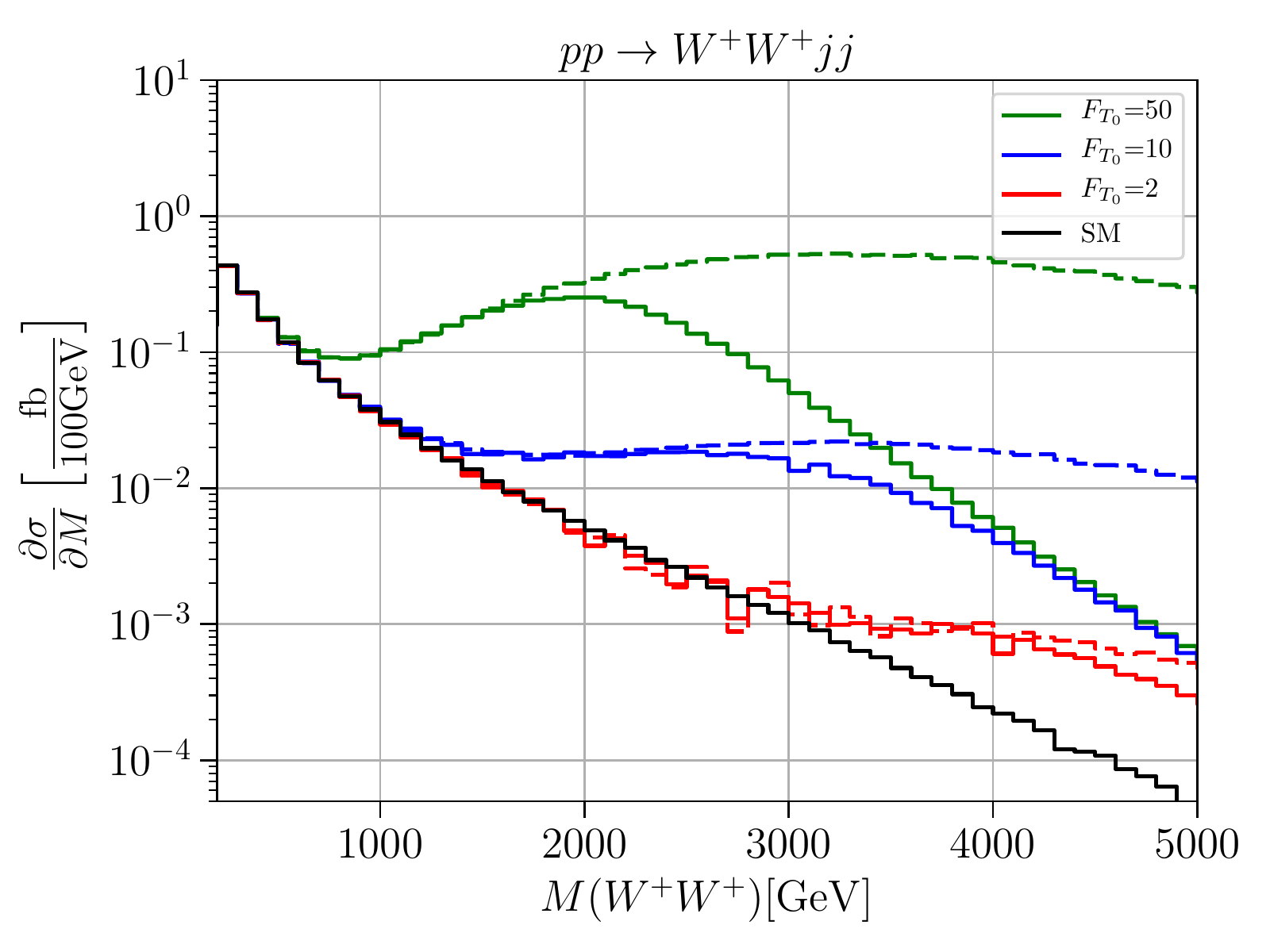}
   \end{subfigure}\hfill%
   \begin{subfigure}[t]{0.5\textwidth}
      \includegraphics[width=\textwidth]{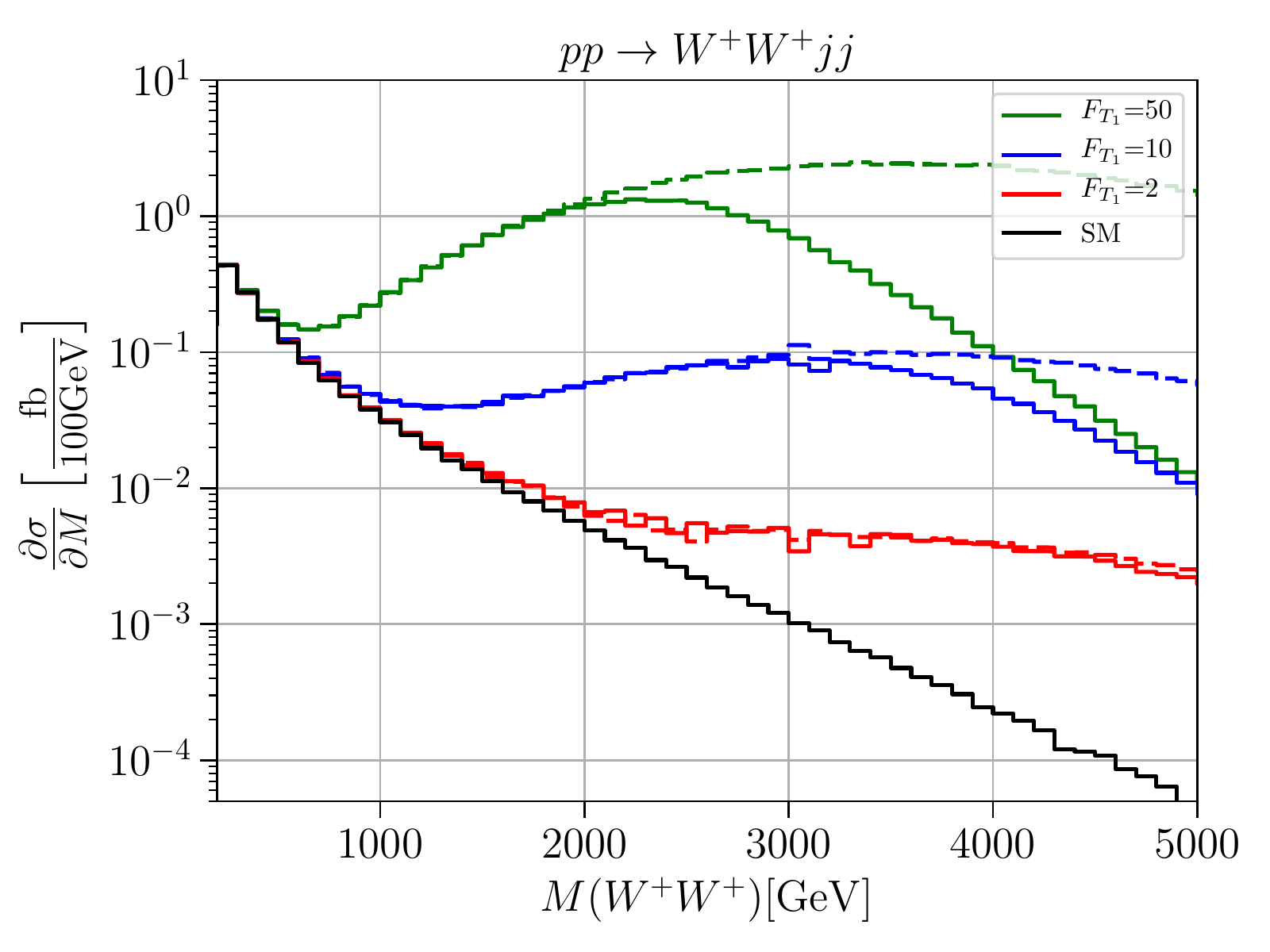}
   \end{subfigure}\\[5pt]%
   \centering
      \begin{subfigure}[t]{0.5\textwidth}
      \includegraphics[width=\textwidth]{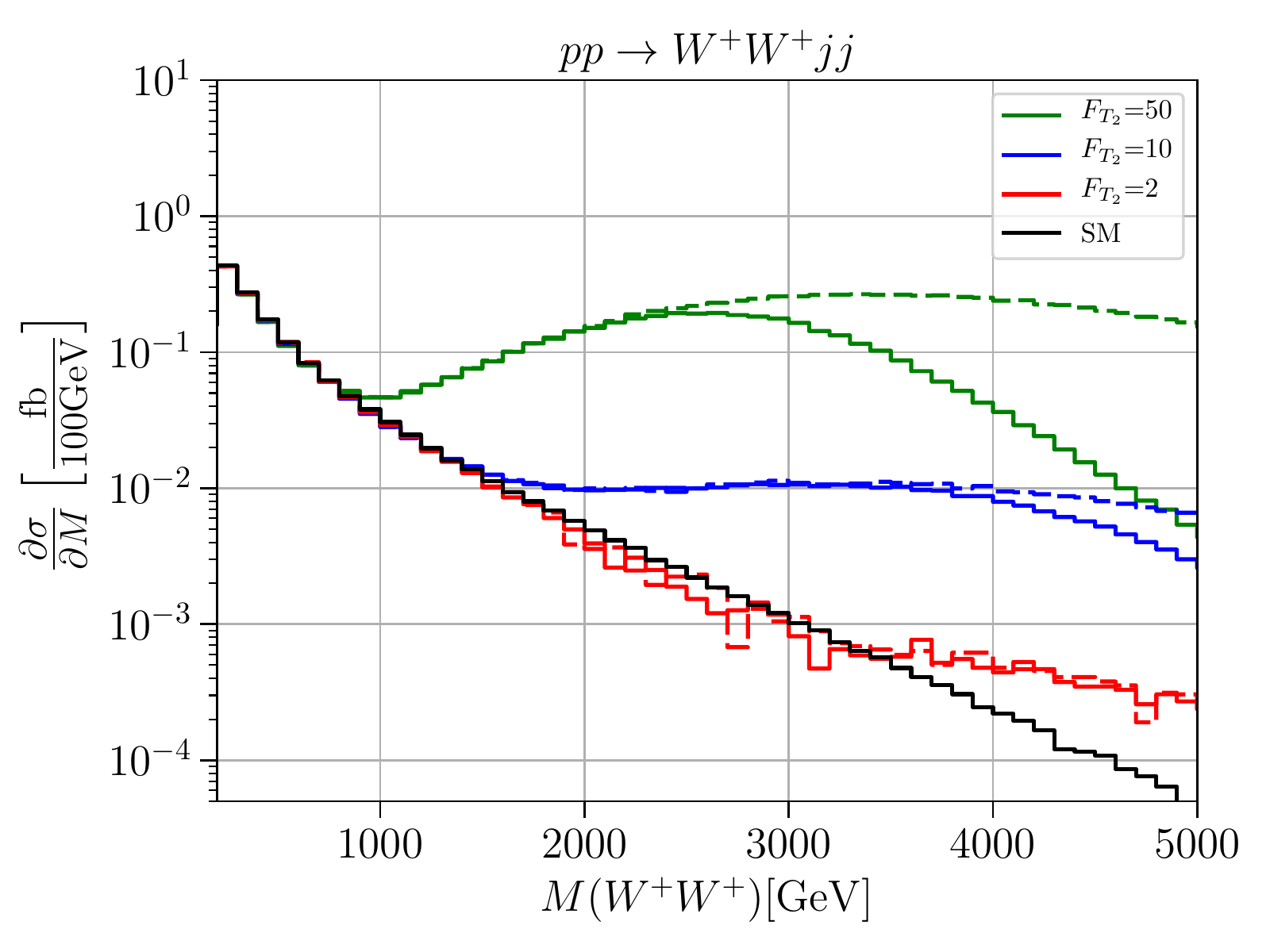}
   \end{subfigure}\\[5pt]%
   \caption{Cross sections differential in the diboson invariant mass
     for the process $pp \to W^+W^+jj$. The solid black line shows the
     SM differential cross section, the green, blue and red lines the
     cross sections with anomalous couplings $F_{T_i}=50\,\
     \TeV^{-4}, F_{T_i}=10\,\TeV^{-4}$ and $F_{T_i}=2\,\TeV^{-4} $ for
     i = 0 (upper left panel), i = 1 (upper right panel), and i = 2
     (lower panel), respectively.
     Solid: unitarized; dashed: naive result.
     Cuts: $M_{jj}>500\,\ \GeV$, $\Delta \eta_{jj}>2.4$,
     $|\eta_j|<4.5$, $p_T^j>20\,\ \GeV$.}
   \label{i:FT_unit} 
\end{figure}

In Fig.~\ref{i:FT_unit} we plot results for the purely transverse
interactions with parameters $F_{T_{0/1/2}}$.  Again, the studied
process is $pp \to W^+W^+ jj$. The $HH$ channel is not affected by these
interactions, 
because the purely transverse 
operators do not contribute to any anomalous coupling involving a Higgs.
We choose three distinct values,
$F_T=2,10,50\;\TeV^{-4}$, with one nonzero coefficient at a time.

In these scenarios, the asymptotic enhancement of the continuum model
over the SM approaches two orders of magnitude.  We may read this
observation as an indication for much larger freedom for new-physics
effects in purely transverse vector-boson interactions, compared to
mixed and purely longitudinal interactions.  This fact should be
accounted for in data analysis.  Nevertheless, also in this class of
models, the naive SMEFT result represented by the dashed lines
overestimates the possible event rates by a large factor.

We emphasize that the above plots, which only indicate the variation
with respect to one of the model parameters, should not be taken
individually as realistic predictions, even if accepting the basic
assumptions regarding a strongly interacting continuum in the
electroweak sector.  They sweep a range of predictions, within the
given model class.  In reality, we expect more than one coefficient to
be present, so a global fit would be required to determine the correct parameter
dependence and the sensitivity of a collider experiment.
On the other hand, we can already conclude that due to the failure of
the naive SMEFT, there is no meaningful description of these processes
that can be viewed as model-independent.  An analysis that compares
actual data to a prediction, apart from the SM result, must choose
among the conceivable (simplified) models for a comparison, of which
we can only show a set of examples.

If several anomalous couplings are present in a model, it is essential to
increase the number of independent observables that enter a global fit of all
parameters.  At the LHC, there is a number of di-boson final state that can be
produced in VBS.  In Appendix~\ref{app-plots} we present results for the
additional VBS final states $W^+W^-$, $W^+Z$, and $ZZ$ which are not as easily
accessible or have smaller leptonic rates compared to $W^+W^+$ but should be
considered in this context, particularly as there are already results
from the LHC experiments for the latter two.  For those results, we
choose a value of $2\;\TeV^{-4}$ for each of the parameters $F_i$,
with only one parameter nonzero at a time.

\section{Simplified Resonance Models}
\label{sec:sim_res}

In this section, we consider simplified models where an anomalous local
interaction resolves into a resonance which saturates a partial-wave
spin-isospin amplitude.  A resonance saturates an elastic channel for finite
energy and exhibits a falloff of the amplitude beyond the peak, before strong
interactions may set in again at higher energy.  This is observed, e.g., for
some isolated bound 
states that precede a strongly 
interacting continuum in QCD.  In Ref.~\cite{Kilian:2015opv}, we described
this class of models in the context of VBS and studied couplings of the
resonance to longitudinal gauge bosons via the scalar
current $J_H^{(4)}$~(\ref{eq:H0-current}).  In
this work, we extend the allowed coupling to transversal bosons.  As an
example, we take a single scalar with a coupling to the current
$J^{(4)}_W$~(\ref{eq:W0-current}).

There are various models of a non-minimal Higgs sector which
effectively lead to a phenomenology of this type.  In general, we
expect couplings of the resonance both to longitudinal and transverse
vector-boson modes.  BSM models which allow a direct coupling of a new
physics particle only to the transverse mode of electroweak gauge
bosons are often very constrained by data
\cite{Buttazzo:2015txu}. Only a few extra-dimensional models
\cite{Giddings:2016sfr,Han:2015cty} including a directly and strongly
coupled spin-2 resonance, for example a KK-graviton, are not as
hampered by experimental data. Other BSM models introduce the
coupling of transverse vector bosons to a new physics particle not
directly, but due to loop contributions. In
Randall-Sundrum~\cite{Randall:1999ee} or ADD
\cite{ArkaniHamed:1998rs} models this could also be achieved
through a top loop~\cite{Geng:2016xin}.

Models with extra scalar resonances typically introduce additional new heavy
particles. For instance, in composite Higgs models the coupling to the
transversal gauge sector can be mediated by
technipions~\cite{Pasechnik:2013bxa} or by heavy
fermions~\cite{Yoon:2017wul,Bauer:2016lbe}.  If the mass scale of such extra
heavy particles is beyond the experimental reach of LHC, the loop
contributions are small and can be parametrized within an EFT.  Effective
couplings of a resonance involve both longitudinal and transversal vector
bosons. In recent
diphoton studies, this EFT framework was also used to estimate the effect of a
possible diphoton resonance~\cite{Buttazzo:2015txu},
\cite{Franceschini:2015kwy,Gupta:2015zzs,Kim:2015ron}. Another class of
models with heavy resonances are Little Higgs
models~\cite{ArkaniHamed:2001nc,ArkaniHamed:2002qy}.  For these models, the
coefficients of the SMEFT as the low-energy expansion have been calculated,
e.g., in Ref.~\cite{Kilian:2003xt}.

In the present paper, we do not refer to a specific scenario.  We
construct a simplified model with transverse couplings of a generic
heavy resonance~$\sigma$.  The effective Lagrangian takes
the following form,
\begin{subequations}
  \label{eq:Lagrangian_scalar-isoscalar}
  \begin{align}
    \LL_{\sigma}&= - \frac{1}{2} \sigma (m_\sigma^2 - \partial^2) \sigma
      + \sigma (J_{\sigma\parallel}+J_{\sigma\perp}) \\
    J_{\sigma\parallel}&=F_{\sigma H}\tr{(\vD_\mu \vH)^\dagger (\vD^\mu \vH)} \\
    J_{\sigma\perp}&= g^2 F_{W\sigma}\sigma \tr{\vW_{\mu\nu}\vW^{\mu\nu}}
      + {g^{\prime}}^2F_{B\sigma}\sigma \tr{\vB_{\mu\nu}\vB^{\mu\nu}}
  \end{align}
\end{subequations}
with three independent coupling parameters.

In the low-energy limit, the scalar resonance can be integrated out,
and we obtain the SMEFT Lagrangian with the following nonzero
coefficients of the dimension-8 operators at
leading order:
\begin{subequations}
  \label{eq:scalarToDim8}
  \begin{align}
    F_{S_0} &= \phantom{-}
    {F_{\sigma H}^2}/{2 m_\sigma^2}
    \\
    F_{M_0} &=
    -{F_{\sigma H}F_{\sigma W}}/{m_\sigma^2}
    \\
    F_{M_2} &=
    -{F_{\sigma H}F_{\sigma B}}/{m_\sigma^2}
    \\
    F_{T_0} &=\phantom{-}
    {F_{\sigma W}^2}/{2m_\sigma^2}
    \\
    F_{T_5} &=\phantom{-}
    {F_{\sigma W}F_{\sigma B}}/{m_\sigma^2}
    \\
    F_{T_8} &= \phantom{-}
    {F_{\sigma B}^2}/{2m_\sigma^2}.
  \end{align}
\end{subequations}
To set the relation between the coupling constant to the electroweak
currents and the resonance mass, we also compute the width of the
scalar resonance:
\begin{align}
  \Gamma (m_\sigma) = \int d\Omega \frac{|\vec{p}|}{32 \pi^2 m_\sigma^2}
  &\left(|\mathcal{M}_{\sigma\rightarrow W^+W^-}|^2
    + \frac{1}{2}|\mathcal{M}_{\sigma\rightarrow ZZ}|^2
    + \frac{1}{2}|\mathcal{M}_{\sigma\rightarrow HH}|^2 \right. \notag\\
    & \left. \quad
    + |\mathcal{M}_{\sigma\rightarrow Z\gamma}|^2
    + \frac{1}{2}|\mathcal{M}_{\sigma\rightarrow \gamma\gamma}|^2
  \right) \, ,
\end{align}
with $|\vec{p}|=m_\sigma/2$. Here, we neglect the masses of the
electroweak gauge bosons in the kinematics of the phase space
vectors. 

The model
with only $F_{\sigma H}$ nonzero has been covered in
Ref.~\cite{Kilian:2015opv}.  For this paper, we set
$F_{\sigma H}=F_{\sigma B}=0$ and keep only $F_{\sigma W}$.  The
resonance width becomes
\begin{align} 
  \label{sigma-width}
  \Gamma_W(m_\sigma)
    &=  \;\frac{3m_\sigma^3}{16 \pi}g^4F_{\sigma W}^2 
    \left ( 1+ \mathcal{O}(1/m^2_\sigma) \right)\, .
\end{align}
The low-energy limit contains only the operator $\mathcal{L}_{T_{0}}$.  We can
thus easily compare distributions with a resonance to the anologous
distributions with a continuum, where both models reduce to the same
low-energy limit.  While the low-energy approximation has a single parameter,
the dimension-eight operator coefficient $F_{T_0}$, the resonance model has
two free parameters, the resonance
mass and the resonance coupling, or alternatively the width.  

\begin{figure}[!ht]
  \begin{center}
    \includegraphics[width=0.6\textwidth]{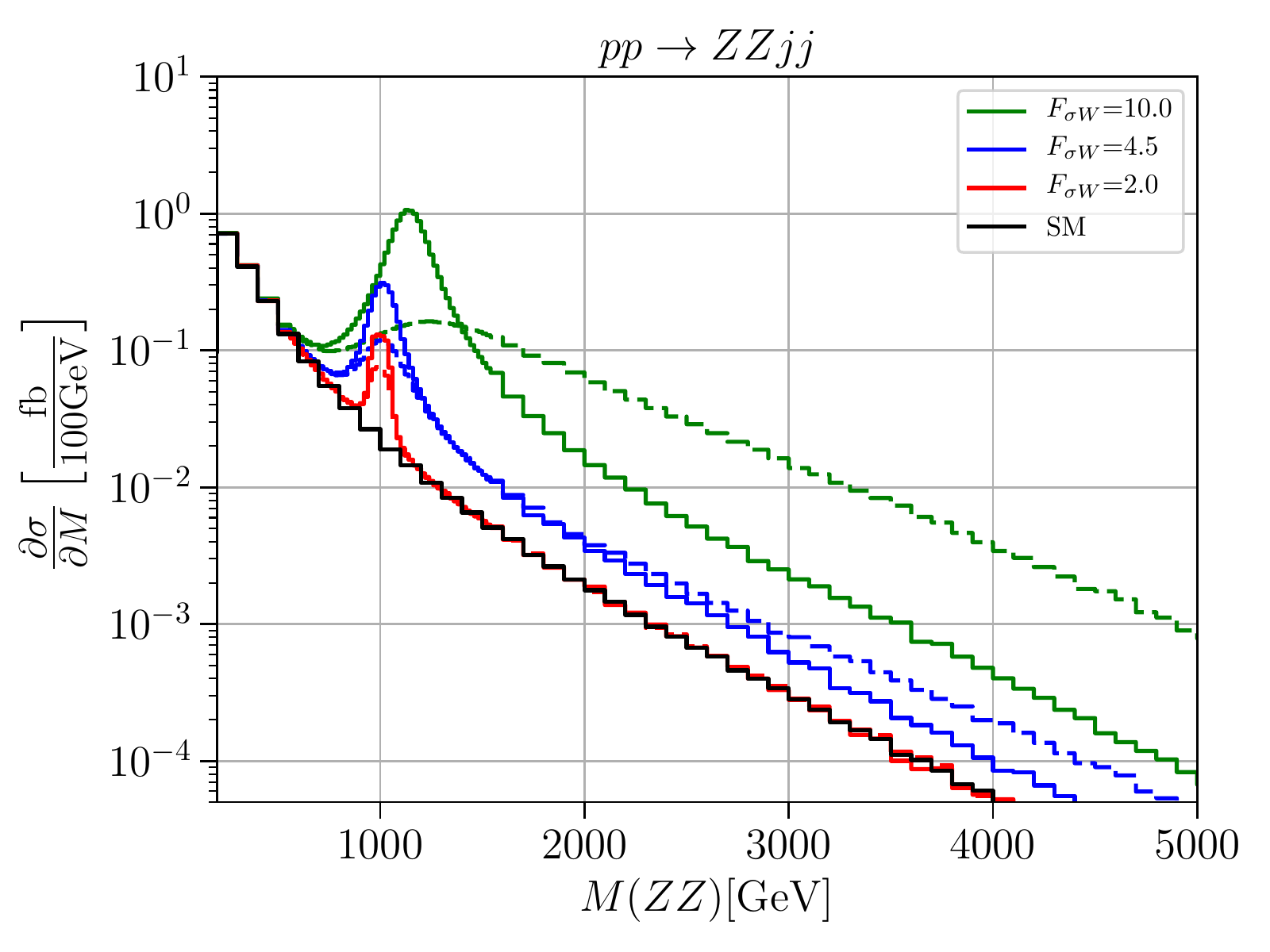}
  \end{center}
  \caption{Cross sections differential in the diboson invariant mass
    for the process $pp \to ZZjj$. The solid black line shows the
    Standard Model differential cross section, the green, blue and red lines the
    cross sections with a scalar resonance with mass $m_\sigma =
    1\,\TeV$  and coupling of $F_{W \sigma}=10.0\,\
    \TeV^{-1}, F_{W \sigma}=4.5\,\TeV^{-1}$ and $F_{W
      \sigma}=2.0\,\TeV^{-1} $, respectively. 
    Solid: unitarized; dashed: naive result.
    Cuts: $M_{jj}>500\,\ \GeV$, $\Delta \eta_{jj}>2.4$,
    $|\eta_j|<4.5$, $p_T^j>20\,\ \GeV$.}
  \label{i:sigma-res}
\end{figure}

We have
implemented the resonance model in the Monte-Carlo generator
\whizard~\cite{Kilian:2007gr}, using the same unitarity projection algorithm
as for the continuum models.  In Fig.~\ref{i:sigma-res}, we show the
invariant-mass distribution of the $ZZ$ final state for a scalar resonance with
mass $m_\sigma= 1 \,\mathrm{TeV}$ and different couplings 
$F_{\sigma W}=10.0,4.5,2.0 \, \TeV^{-1}$.
These values correspond to the anomalous quartic
coupling $F_{T_0}=50,10,2 \, \TeV^{-4}$ if the scalar resonance is integrated
out.  The dashed lines show the naive result of implementing the scalar
resonance as an extra particle with its width given by the
formula~(\ref{sigma-width}).  The solid lines show the unitary projection for
each coupling value, respectively.

This plot illustrates two properties of resonance models.  First of all, we
observe that the unitary projection has two effects: on the resonance, the
peak becomes narrower and more pronounced.  This is the result of subleading
terms in the width formula, which we did not include in the naive result but
which are accounted for by the unitary projection.  Asymptotically, the
amplitude is suppressed by the projection.  This is the result of saturating
partial waves by $s^2$ terms which originate from the derivative coupling.

Since a derivative coupling is a typical feature of strong interactions where
couplings involve form factors, and a necessary property of resonances with
higher spin, the asymptotic effect of unitarity saturation is essential for a
complete description.  The T-matrix projection is a method for implementing
unitarity in the model for the whole kinematical range.

\begin{figure}[!ht]
  \begin{center}
    \includegraphics[width=0.6\textwidth]{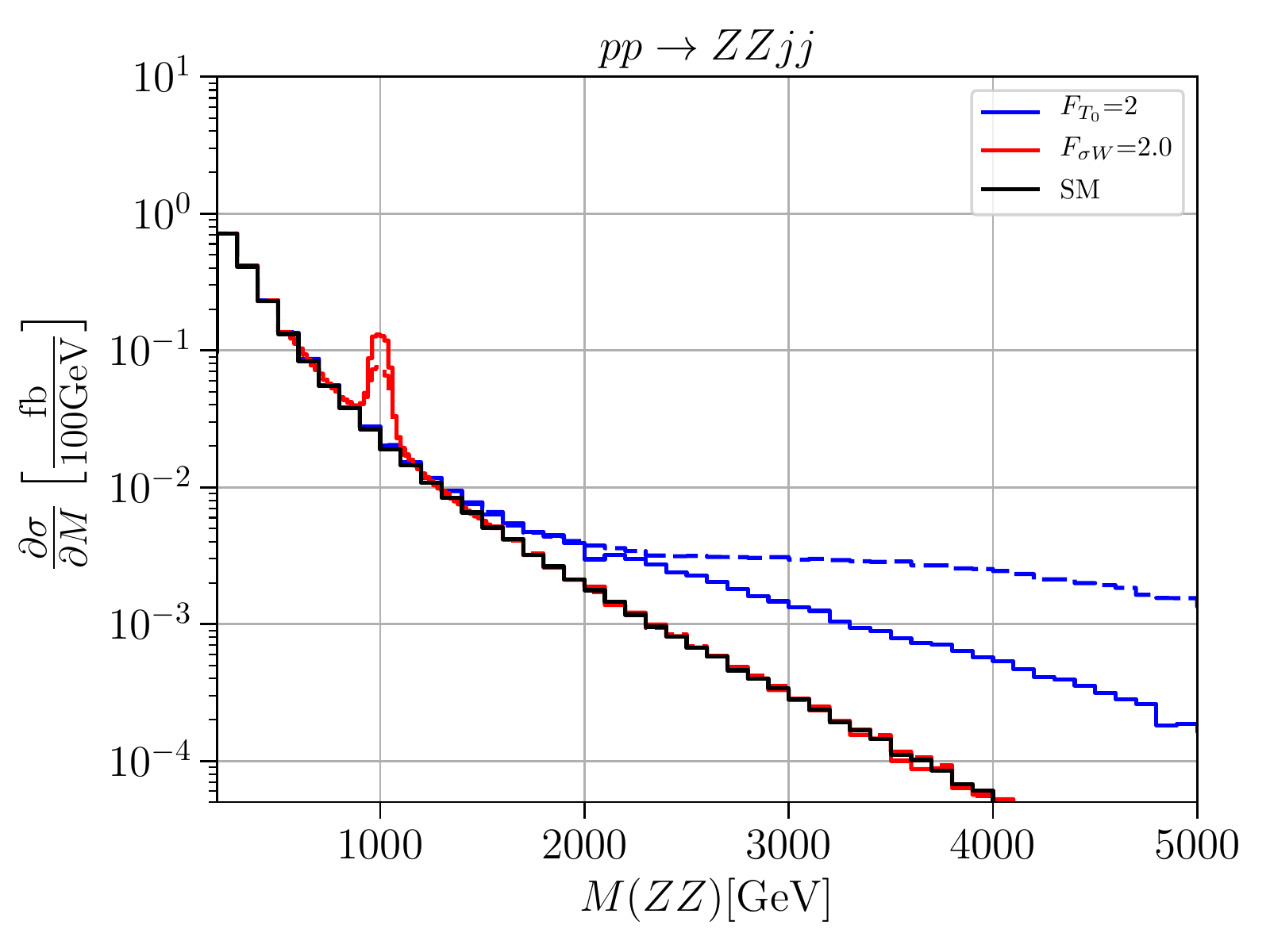}
  \end{center}
  \caption{Cross sections differential in the diboson invariant mass
    for the process $pp \to ZZjj$. The solid black line shows the
    Standard Model, the blue lines show
    an anomalous coupling  $F_{T_i}=2\,\TeV^{-4} $
    and the red lines show a scalar resonance with mass
     $m_\sigma = 1\,\TeV$ and coupling of $F_{\sigma W}=2.0\,~\TeV^{-2}$.
    Solid: unitarized; dashed: naive result.
    Cuts: $M_{jj}>500\,\ \GeV$, $\Delta \eta_{jj}>2.4$,
    $|\eta_j|<4.5$, $p_T^j>20\,\ \GeV$.}
  \label{i:sigma-res-eft}
\end{figure}

In Fig.~\ref{i:sigma-res-eft}, we compare the simplified model with a
scalar resonance with mass $m_\sigma= 1 \,\mathrm{TeV}$ and coupling $F_{\sigma
W}=2.0 \, \TeV^{-1}$ (red) to the corresponding EFT result with the matching
anomalous quartic coupling $F_{T}=2 \,\TeV^{-4}$ (blue), with and without
unitary projection (solid vs.\ dashed).   This is a rather small coupling, and
the resonance behaves almost like an elementary particle.  The peak is not
approximated at all by the EFT operator description.  We may argue that for
such a type of model, the EFT is useful only in the case of strong couplings
and broad resonances.  We also see that the high-energy behavior of the EFT
approximation has no resemblance to the high-energy behavior of the resonance
model, regardless of unitarity projection.  

We conclude that including resonance models in the description allows us to
smoothly interpolate between weakly and strongly interacting models.  This
interpolation may leave the applicability range of perturbative expansions,
but does not require to deal with unphysical behavior as a calculational
artefact.


 

\section{Implications for LHC Analyses and Conclusions}
\label{sec:results}

The ATLAS and CMS experiments have analyzed the early-stage LHC data
with respect to the sensitivity to VBS parameters.  Table \ref{table:exp_limits}
summarizes published results, expressed in terms of the unmodified
SMEFT parameterization with dimension-eight operators included.

\begin{table}[!th]
  \begin{center}  
  \begin{tabular}{c| c c c c}
   Coefficient &CMS\&ATLAS \cite{Green:2016trm} 
   &ATLAS\cite{Aaboud:2016uuk,Sekulla:2016yku} &
   CMS \cite{Sirunyan:2017ret,Sirunyan:2017fvv}  &CMS reweighted \\
   {}[TeV$^{-4}$] &8 TeV, EFT & 8 TeV, T-matrix &
    13 TeV, EFT  & 13 TeV, EFT \\
  \hline  \\
  $f_{S_0}/\Lambda^4$&[-38,40] & & [-7.7,7.7]  &[-7.7,7.7]                                   \\
  $f_{S_1}/\Lambda^4$&[-118,120] & & [-21.6,21.8]&[-21.6,21.8]             \\
  $F_{S_0}$ & [-70,70]& [-104,130] & \\
  $F_{S_1}$ & [-118,120] &[-122,144] &   \\
  $f_{M_0}/\Lambda^4$&[-18,18] & & [-6.0,5.9]  &[-13.8,14.1]                           \\
  $f_{M_1}/\Lambda^4$&[-44,47] & & [-8.7,9.1]  &[-21.4,20.4]                           \\
  $f_{M_6}/\Lambda^4$&[-65,63] & & [-11.9,11.8]&[-27.7,27.9]                           \\
  $f_{M_7}/\Lambda^4$&[-70,66] & & [-13.3,12.9]&[-30.3,31.2]                          \\
  $f_{T_0}/\Lambda^4$&[-4.2,4.6] & & [-0.46,0.44]&[-2.53,2.42]                          \\
  $f_{T_1}/\Lambda^4$&[1.9,2.2]  & & [-0.28,0.31]&[-1.54,1.71]                           \\
  $f_{T_2}/\Lambda^4$&[-5.2,6.4]  & & [-0.89,1.02]&[-4.9,5.6]                          \\
  $f_{T_9}/\Lambda^4$ &[-6.9,6.9] & & [-1.8,1.8]&[-7.5,7.5]                          \\
  \end{tabular}
  \end{center}
    \caption{ Observed Limits of ATLAS and CMS of complete LHC data at
      $\sqrt{s}=8 \mathrm{TeV}$ and current observed limits of CMS at
      $\sqrt{s}=13 \mathrm{TeV}$ using the naive EFT model 
      and the T-matrix model. The last column show the limits in natural
      reweighting of field strength tensors: $\vW^{\mu\nu}\rightarrow \ii g
      \vW^{\mu\nu}$, $\vB^{\mu\nu}\rightarrow \ii g^\prime \vB^{\mu\nu}$.}
  \label{table:exp_limits}
\end{table}
 
In view of the results presented in the preceding sections, we have to discuss
the physical relevance of the published exclusion bounds.  In principle, the
SMEFT approach provides a well-defined framework.  However, our findings
confirm the expectation that the SMEFT expansion, applied to VBS as a LHC
process, does not provide a systematic expansion or meaningful description of
the complete data set.  For nonvanishing dimension-eight coefficients, the
amplitudes rise steeply with energy, such that a problem invariably arises
within the accessible kinematic range.  This happens for \emph{any} set of
parameter values, unless the dimension-eight coefficients are so small that
the prediction remains entirely indistinguishable from the SM.

The measurements acquire a physical interpretation only within the context of
a unitary model.  For instance, we may apply a straightforward T-matrix
projection to the naive extrapolation and thus consider a unitary simplified
model that is smoothly matched to the low-energy SMEFT, depending on the same
parameters that in the low-energy act as dimension-eight operator
coefficients.  We find that the sensitivity of this unitary model to the free
parameters is much weaker than the naive calculation would suggest, likely by
an order of magnitude.  Since the minimal T-matrix projection interpolates the
low-energy behavior with asymptotic saturation of the elastic channel, this
particular projection provides us with the ultimate limitation to the
achievable parameter sensitivity.

We conclude that any such description or theoretical prediction of non-SM
behavior has to depart from the model-independent paradigm.  Otherwise, data
analysis has to artificially remove kinematical regions from the data sample,
losing valuable information.  A well-defined universal but model-dependent
parameterization is certainly possible, however, without losing contact to the
SMEFT as a systematic description of the low-energy region.

In this work, we have demonstrated the construction of unitary projections
that yield usable simplified models for otherwise unknown new physics.
Extending previous work, we have included transverse vector-boson
polarization modes together with final-state Higgs bosons in the completed
framework.  None of our models is UV complete or otherwise meaningful as a
prediction.  However, for the purpose of estimating the prospects for future
measurements in quantitative terms, such a set of simplified models becomes a
useful tool.  Applying the direct T-matrix projection to the straightforward
extrapolation of the SMEFT amplitudes with dimension-eight operators, we
obtain a natural interpolation between the low-energy range which is well
understood, and high-energy amplitudes which saturate the unitarity limits.
This sets the scale for more refined models, such as the model of a singlet
scalar coupled to transverse gauge bosons which we also have considered in
some detail.  In essence, we obtain parameter-dependent upper limits for event
rates of all processes for all energy ranges, which refined models have to
respect.

The lesson to be learned from such results is twofold.  Firstly, we read off
the range of event rates and distributions that we can possibly expect from
LHC experiments, for any underlying model.  This range can only be exceeded if
some natural, basic assumptions are violated by Nature.  More precisely,
violations of the assumptions would point to (a) fermions directly involved in
new (strong) Higgs-sector interactions, or (b) gauge symmetry being just a
low-energy accident, or (c) four-dimensional quantum field theory becoming
invalid.  Either scenario appears to be unlikely given the success of the SM
in describing low-energy data, in particular in the flavor sector.  For this
reason, we believe that the quantitative results obtained within the framework
of unitary simplified models reliably exhaust the range that can be expected
from real data.

Secondly, the framework of unitary projection, now extended to transverse
polarizations, enables any theoretical idea or model of the Higgs sector as a
viable model for Monte-Carlo simulations.  I.e., the projection satisfies the
applicable unitarity constraints, correctly couples to fermionic currents, and
the collider environment is described in consistency with the
analogous SM calculation.  In short, the model can be compared to data without
further approximations or simplifications.  The downside is
that for VBS processes, there is no usable model-independent framework, and
any study has to agree on a particular model class and assumptions for
interpreting the results.  Furthermore, the arbitrariness in the
parameterization mandates the inclusion of all quantum-number combinations and
global fits, which would greatly benefit from a larger set of observables such
as can be obtained at high-energy lepton colliders supplementing the LHC.

Finally, we add a remark on Higgs pair-production.  This final state has
received particular attention since it is sensitive to the triple-Higgs
coupling and thus to the Higgs potential.  Higgs
pairs can result from gluon or massive vector-boson fusion.  The latter
channel has the particular feature of extra taggable forward jets.  In a
generic EFT description, the Higgs pair-production process in VBF receives
various contributions that can be attributed to higher-dimensional operators,
and the Higgs potential correction is only one of those.  Furthermore, our
results show that dimension-eight operators can drastically enhance the Higgs
pair-production rate by three orders of magnitude before unitarity limits set
in.  Since there is no reason to expect the operator series expansion to stop
at dimension six, we are forced to argue that any analysis of Higgs
pair-production data that confines itself to a truncated expansion has to be
taken with a grain of salt.  On the other hand, linear gauge invariance
relates anomalous effects in Higgs pair-production to anomalous effects in VBS
at the same order.  Future LHC Higgs pair-production analyses thus should
correlate all accessible boson-production channels.  The interpretation,
however, will rely on model-dependent approaches such as the one that we
present in this paper.

\subsection*{Acknowledgments}

C.F.\ has been supported by a HYT Fellowship of the University of Siegen.
W.K.\ thanks the CLICdp and Theory Groups for their hospitality at CERN,
where part of this work was completed.  We also acknowledge valuable
discussions with C.~Grojean, M.-A. Pleier, M.~Rauch, M. Szleper and
A.~Wulzer, and we thank D.~Zeppenfeld for carefully reading and useful
comments on the manuscript. J.R.R. acknowledges support by the EU COST
Network "VBScan" (COST Action CA16108). 


\appendix

\clearpage
\section{Isospin-spin amplitudes}

In this section we collect the different isospin-spin eigenamplitudes
for the different combinations of helicities of the electroweak gauge
bosons according to decomposition into Wigner functions in
Eq.~\ref{e:isospinspincalc}. As explained in the last paragraph of
Sec.~\ref{sec:currents}, we take here only the Wilson coefficients of
those transversal and mixed operators $\mathcal{L}_{T_i}$ and
$\mathcal{L}_{M_i}$, respectively, with indices $i=0,1,2$ into account
for which there is custodial $SU(2)_C$ conserved. 

Already in Ref.~\cite{Kilian:2015opv} it was shown that the kinematic
functions for the unitarized amplitudes for resonances are not simple
powers of $s$, but contain logarithms and pole-like rational functions
of $s$ in general. For the isospin-spin amplitudes in the case of a
isoscalar scalar resonance in Table~\ref{t:isospin-spin-coeff_res}, we
define the following kinematic functions:

\begin{align}
X(s,m)&=\frac{3s^2}{s-m^2} \\
\mathcal{S}_0(s,m)&=2m^2+2\frac{m^4}{s}
\log \left(\frac{m^2}{s+m^2} \right) -s\\
\mathcal{S}_1(s,m)&=4\frac{m^4}{s}+6m^4\left(2m^2+s\right)
\log \left(\frac{m^2}{s+m^2} \right) +\frac{s}{3}\\
\mathcal{S}_2(s,m)&=6\frac{m^4}{s^2}\left(2m^2+s\right)+2\frac{m^4}{s^3}
\left(6m^4+6m^2 s+s^2\right) \log \left(\frac{m^2}{s+m^2} \right) \\
\tilde{\mathcal{S}_2}(s,m)&=\frac{m^2}{3}-\frac{m^4}{2s}
+\frac{m^6}{s^2} + \frac{m^8}{s^3}\log \left(\frac{m^2}{s+m^2} \right)
-\frac{s}{4}
\end{align}

\begin{table}[hbtp]
\centering
\begin{tabular}{c|ccc|ccc|ccc|c}
\diagbox{i}{j} &   & 0 &   &   & 1 &   &   & 2 &   &  $\boldsymbol{\lambda}$    \\
     \hline
  & $\frac{3}{2}$ & $-\frac{3}{8}$ & $\frac{3}{16}$ & 0 & 0 & 0 & 0 & 0 & 0 & \begin{tabular}{cccc}
$+$ & $+$ & $0$ & $0$ \\
\end{tabular} \\
  & 0 & 0 & 0 & 0 & 0 & 0 & 0 & $\frac{1}{20}\sqrt{\frac{3}{2}}$ & $\frac{1}{40}\sqrt{\frac{3}{2}}$ & \begin{tabular}{cccc}
$+$ & $-$ & $0$ & $0$ \\
\end{tabular} \\
0 & 0 & 0 & 0 & $-\frac{1}{8}$ & $\frac{1}{32}$ & $\frac{7}{192}$ & $\frac{1}{40}$ & $-\frac{1}{160}$ & $\frac{3}{320}$ & \begin{tabular}{cccc}
$+$ & $0$ & $-$ & $0$ \\
\end{tabular} \\
  & 0 & 0 & 0 & $\frac{1}{8}$ & $-\frac{1}{32}$ & $-\frac{7}{192}$ & $\frac{1}{40}$ & $-\frac{1}{160}$ & $\frac{3}{320}$ & \begin{tabular}{cccc}
$+$ & $0$ & $0$ & $-$ \\
\end{tabular} \\
  & 0 & 0 & 0 & 0 & $\frac{1}{12}$ & $\frac{1}{24}$ & 0 & 0 & 0 & \begin{tabular}{cccc}
$+$ & $0$ & $+$ & $0$ \\
\end{tabular} \\
  & 0 & 0 & 0 & 0 & $-\frac{1}{12}$ & $-\frac{1}{24}$ & 0 & 0 & 0 & \begin{tabular}{cccc}
$+$ & $0$ & $0$ & $+$ \\
\end{tabular} \\
       \hline
  & 0 & 0 & 0 & 0 & 0 & $\frac{1}{24}$ & 0 & 0 & 0 & \begin{tabular}{cccc}
$+$ & $+$ & $0$ & $0$ \\
\end{tabular} \\
  & 0 & 0 & 0 & 0 & 0 & 0 & 0 & 0 & 0 & \begin{tabular}{cccc}
$+$ & $-$ & $0$ & $0$ \\
\end{tabular} \\
1 & 0 & 0 & 0 & $-\frac{1}{8}$ & $\frac{1}{32}$ & $\frac{1}{96}$ & $\frac{1}{40}$ & $-\frac{1}{160}$ & $\frac{1}{160}$ & \begin{tabular}{cccc}
$+$ & $0$ & $-$ & $0$ \\
\end{tabular} \\
  & 0 & 0 & 0 & $-\frac{1}{8}$ & $\frac{1}{32}$ & $\frac{1}{96}$ & $-\frac{1}{40}$ & $\frac{1}{160}$ & $-\frac{1}{160}$ & \begin{tabular}{cccc}
$+$ & $0$ & $0$ & $-$ \\
\end{tabular} \\
  & 0 & 0 & 0 & 0 & $\frac{1}{12}$ & $\frac{1}{24}$ & 0 & 0 & 0 & \begin{tabular}{cccc}
$+$ & $0$ & $+$ & $0$ \\
\end{tabular} \\
  & 0 & 0 & 0 & 0 & $\frac{1}{12}$ & $\frac{1}{24}$ & 0 & 0 & 0 & \begin{tabular}{cccc}
$+$ & $0$ & $0$ & $+$ \\
\end{tabular} \\
       \hline
  & 0 & 0 & 0 & 0 & 0 & 0 & 0 & 0 & 0 & \begin{tabular}{cccc}
$+$ & $+$ & $0$ & $0$ \\
\end{tabular} \\
  & 0 & 0 & 0 & 0 & 0 & 0 & 0 & 0 & 0 & \begin{tabular}{cccc}
$+$ & $-$ & $0$ & $0$ \\
\end{tabular} \\
2 & 0 & 0 & 0 & $-\frac{1}{8}$ & $\frac{1}{32}$ & $-\frac{1}{24}$ & $\frac{1}{40}$ & $-\frac{1}{160}$ & 0 & \begin{tabular}{cccc}
$+$ & $0$ & $-$ & $0$ \\
\end{tabular} \\
  & 0 & 0 & 0 & $\frac{1}{8}$ & $-\frac{1}{32}$ & $\frac{1}{24}$  & $\frac{1}{40}$ & $-\frac{1}{160}$ & 0 & \begin{tabular}{cccc}
$+$ & $0$ & $0$ & $-$ \\
\end{tabular} \\
  & 0 & 0 & 0 & 0 & $\frac{1}{12}$ & $\frac{1}{24}$ & 0 & 0 & 0 & \begin{tabular}{cccc}
$+$ & $0$ & $+$ & $0$ \\
\end{tabular} \\
  & 0 & 0 & 0 & 0 & $-\frac{1}{12}$ & $-\frac{1}{24}$ & 0 & 0 & 0 & \begin{tabular}{cccc}
$+$ & $0$ & $0$ & $+$ \\
\end{tabular} \\
       \hline
  & $c_0$ & $c_1$ & $c_2$ & $c_0$ & $c_1$ & $c_2$ & $c_0$ & $c_1$ & $c_2$ &
\end{tabular}
\caption{Coefficients of the isospin-spin amplitudes calculated with
  eq.~(\ref{e:isospinspincalc}) for the mixed operators
  $\mathcal{L}_{M_i}$ depending on the helicity of the incoming and
  outgoing particles. The isospin spin amplitudes are given by
  $A_{ij}(s; \boldsymbol{\lambda})=(c_0 F_{M_0} + c_1 F_{M_1} + c_2
  F_{M_7})g^2 s^2$.} 
\label{t:isospin-spin-coeff_m}
\end{table}

\begin{table}[hbtp]
\centering
\begin{tabular}{c|ccc|ccc|ccc|c}
\diagbox{i}{j} &   & 0 &   &   & 1 &   &   & 2 &   &  $\boldsymbol{\lambda}$    \\
     \hline
  & -6 & -2 & $-\frac{5}{2}$ & 0 & 0 & 0 & 0 & 0 & 0 & \begin{tabular}{cccc}
$+$ & $+$ & $+$ & $+$ \\
\end{tabular} \\
0 & 0 & 0 & 0 & 0 & 0 & 0 & $-\frac{2}{5}$ & $-\frac{4}{5}$ & $-\frac{1}{2}$ & \begin{tabular}{cccc}
$+$ & $-$ & $+$ & $-$ \\
\end{tabular} \\
  & 0 & 0 & 0 & 0 & 0 & 0 & $-\frac{2}{5}$ & $-\frac{4}{5}$ & $-\frac{1}{2}$ & \begin{tabular}{cccc}
$+$ & $-$ & $-$ & $+$ \\
\end{tabular} \\
  & $-\frac{22}{3}$ & $-\frac{14}{3}$ & $-\frac{11}{6}$ & 0 & 0 & 0 & $-\frac{2}{15}$ & $-\frac{4}{15}$ & $-\frac{1}{30}$ & \begin{tabular}{cccc}
$+$ & $+$ & $-$ & $-$ \\
\end{tabular} \\
       \hline
  & 0 & 0 & 0 & 0 & 0 & 0 & 0 & 0 & 0 & \begin{tabular}{cccc}
$+$ & $+$ & $+$ & $+$ \\
\end{tabular} \\
1 & 0 & 0 & 0 & 0 & 0 & 0 & $\frac{2}{5}$ & $-\frac{1}{5}$ & 0 & \begin{tabular}{cccc}
$+$ & $-$ & $+$ & $-$ \\
\end{tabular} \\
  & 0 & 0 & 0 & 0 & 0 & 0 & $-\frac{2}{5}$ & $\frac{1}{5}$ & 0 & \begin{tabular}{cccc}
$+$ & $-$ & $-$ & $+$ \\
\end{tabular} \\
  & 0 & 0 & 0 & $\frac{2}{3}$ & $-\frac{1}{3}$ & $\frac{1}{6}$ & 0 & 0 & 0 & \begin{tabular}{cccc}
$+$ & $+$ & $-$ & $-$ \\
\end{tabular} \\
       \hline
  & 0 & -2 &-1 & 0 & 0 & 0 & 0 & 0 & 0 & \begin{tabular}{cccc}
$+$ & $+$ & $+$ & $+$ \\
\end{tabular} \\
2 & 0 & 0 & 0 & 0 & 0 & 0 & $-\frac{2}{5}$ & $-\frac{1}{5}$ & $-\frac{1}{5}$ & \begin{tabular}{cccc}
$+$ & $-$ & $+$ & $-$ \\
\end{tabular} \\
  & 0 & 0 & 0 & 0 & 0 & 0 & $-\frac{2}{5}$ & $-\frac{1}{5}$ & $-\frac{1}{5}$ & \begin{tabular}{cccc}
$+$ & $-$ & $-$ & $+$ \\
\end{tabular} \\
  & $-\frac{4}{3}$ & $-\frac{8}{3}$ & $-\frac{1}{3}$ & 0 & 0 & 0 & $-\frac{2}{15}$ & $-\frac{1}{15}$ & $-\frac{1}{30}$ & \begin{tabular}{cccc}
$+$ & $+$ & $-$ & $-$ \\
\end{tabular} \\
       \hline
  & $c_0$ & $c_1$ & $c_2$ & $c_0$ & $c_1$ & $c_2$ & $c_0$ & $c_1$ & $c_2$ &
\end{tabular}
\caption{Coefficients of the isospin-spin amplitudes calculated with
  eq.~(\ref{e:isospinspincalc}) for the transversal operators
  $\mathcal{L}_{T_i}$ depending on the helicity of the incoming and
  outgoing particles. The isospin-spin amplitudes are given by
  $A_{ij}(s; \boldsymbol{\lambda})=(c_0 F_{T_0} + c_1 F_{T_1} + c_2
  F_{T_2})g^4 s^2$.} 
\label{t:isospin-spin-coeff_t}
\end{table}


\begin{table}[hbtp]
\centering
\begin{tabular}{c|c|c|c|c}
\diagbox{i}{j} & 0 & 1 & 2 & $\boldsymbol{\lambda}$ \\
\hline
  & $X(s,m)$ & 0 & 0 & \begin{tabular}{cccc}
$+$ & $+$ & $+$ & $+$ \\
\end{tabular} \\
0 & 0 & 0 & $\tilde{\mathcal{S}_2}(s,m)$ & \begin{tabular}{cccc}
$+$ & $-$ & $+$ & $-$ \\
\end{tabular} \\
  & 0 & 0 & $\tilde{\mathcal{S}_2}(s,m)$ & \begin{tabular}{cccc}
$+$ & $-$ & $-$ & $+$ \\
\end{tabular} \\
  & $X(s,m)+\mathcal{S}_0(s,m)$ & 0 & $\mathcal{S}_2(s,m)$ & \begin{tabular}{cccc}
$+$ & $+$ & $-$ & $-$ \\
\end{tabular} \\
\hline
  & 0 & 0 & 0 & \begin{tabular}{cccc}
$+$ & $+$ & $+$ & $+$ \\
\end{tabular} \\
1 & 0 & 0 & $-\tilde{\mathcal{S}_2}(s,m)$ & \begin{tabular}{cccc}
$+$ & $-$ & $+$ & $-$ \\
\end{tabular} \\
  & 0 & 0 & $\tilde{\mathcal{S}_2}(s,m)$ & \begin{tabular}{cccc}
$+$ & $+$ & $+$ & $+$ \\
\end{tabular} \\
  & 0 & $\mathcal{S}_1(s,m)$ & 0 & \begin{tabular}{cccc}
$+$ & $+$ & $-$ & $-$ \\
\end{tabular} \\
\hline
  & 0 & 0 & 0 & \begin{tabular}{cccc}
$+$ & $+$ & $+$ & $+$ \\
\end{tabular} \\
2 & 0 & 0 & $\tilde{\mathcal{S}_2}(s,m)$ & \begin{tabular}{cccc}
$+$ & $-$ & $+$ & $-$ \\
\end{tabular} \\
  & 0 & 0 & $\tilde{\mathcal{S}_2}(s,m)$ & \begin{tabular}{cccc}
$+$ & $-$ & $-$ & $+$ \\
\end{tabular} \\
  & $\mathcal{S}_0(s,m)$ & 0 & $\mathcal{S}_2(s,m)$ & \begin{tabular}{cccc}
$+$ & $+$ & $-$ & $-$ \\
\end{tabular} \\
\end{tabular}
\caption{Coefficients $c$ of the isospin spin amplitudes calculated with eq.~(\ref{e:isospinspincalc}) for the isoscalar scalar resonance $\mathcal{L}_{\sigma W}$ depending on the helicity of the incoming and outgoing particles. The isospin spin amplitudes are given by $A_{ij}(s; \boldsymbol{\lambda})=c g^4 s^2$.}
\label{t:isospin-spin-coeff_res}
\end{table} 

\clearpage
\section{Additional Numerical Results}
\label{app-plots}

In this section, we display results for the invariant-distribution of the LHC
processes $pp\to W^+W^-jj$, $ZZjj$, and $W^+Zjj$ which supplement the results
for the $W^+W^+$ and $HH$ channels in the main text.  For all processes, we
present the SM distribution together with the corresponding distribution of
the continuum simplified model, one free parameter varied at a time, with a
universal parameter value of $2\;\TeV^{-4}$.

\begin{figure}
  \begin{subfigure}{0.5\textwidth}
    \includegraphics[width=0.9 \linewidth]{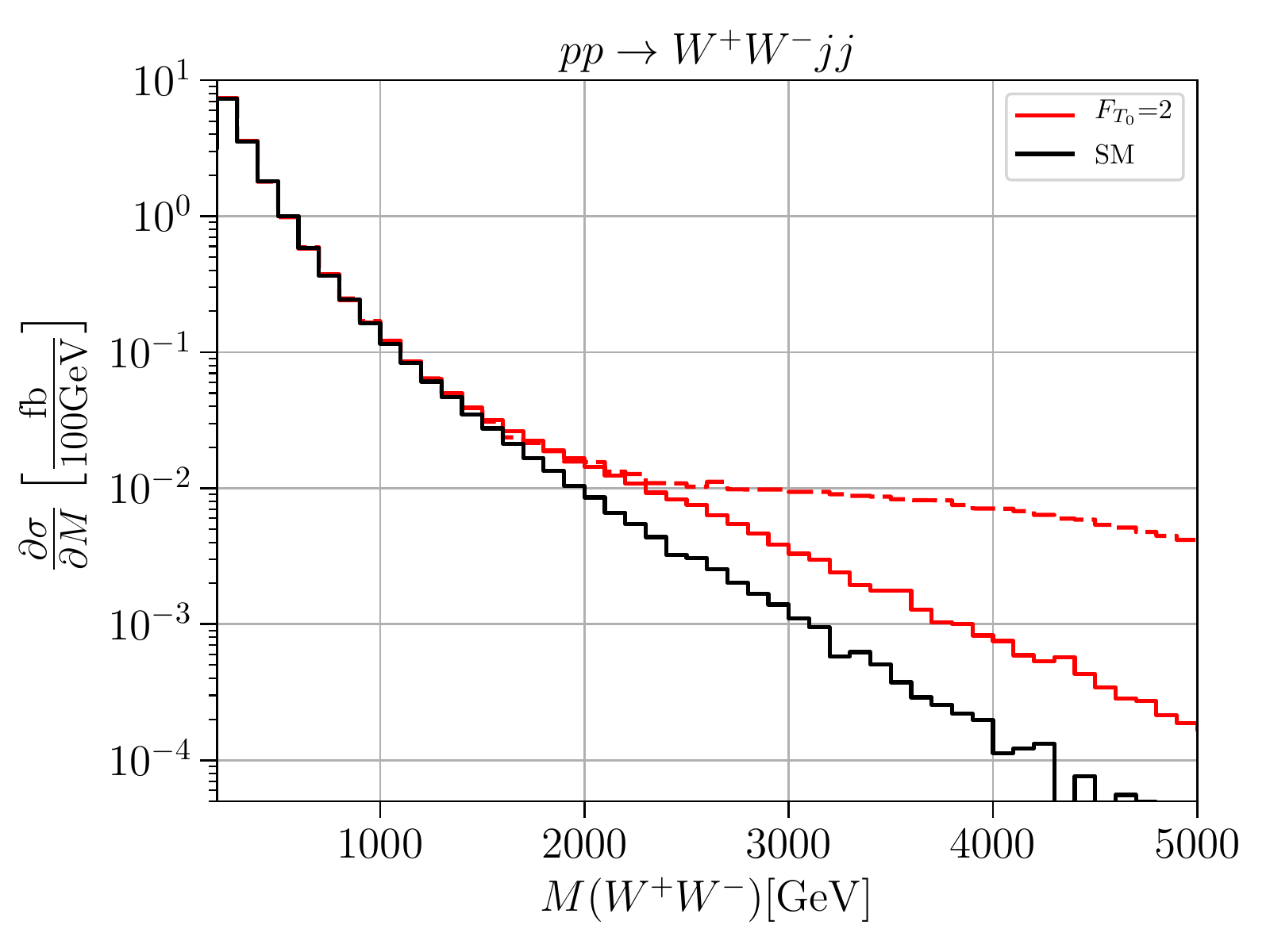}
    \caption{$\mathcal{O}_{T_0}$}
    \label{fig:WpWm-t0}
  \end{subfigure}
  \begin{subfigure}{0.5\textwidth}
    \includegraphics[width=0.9 \linewidth]{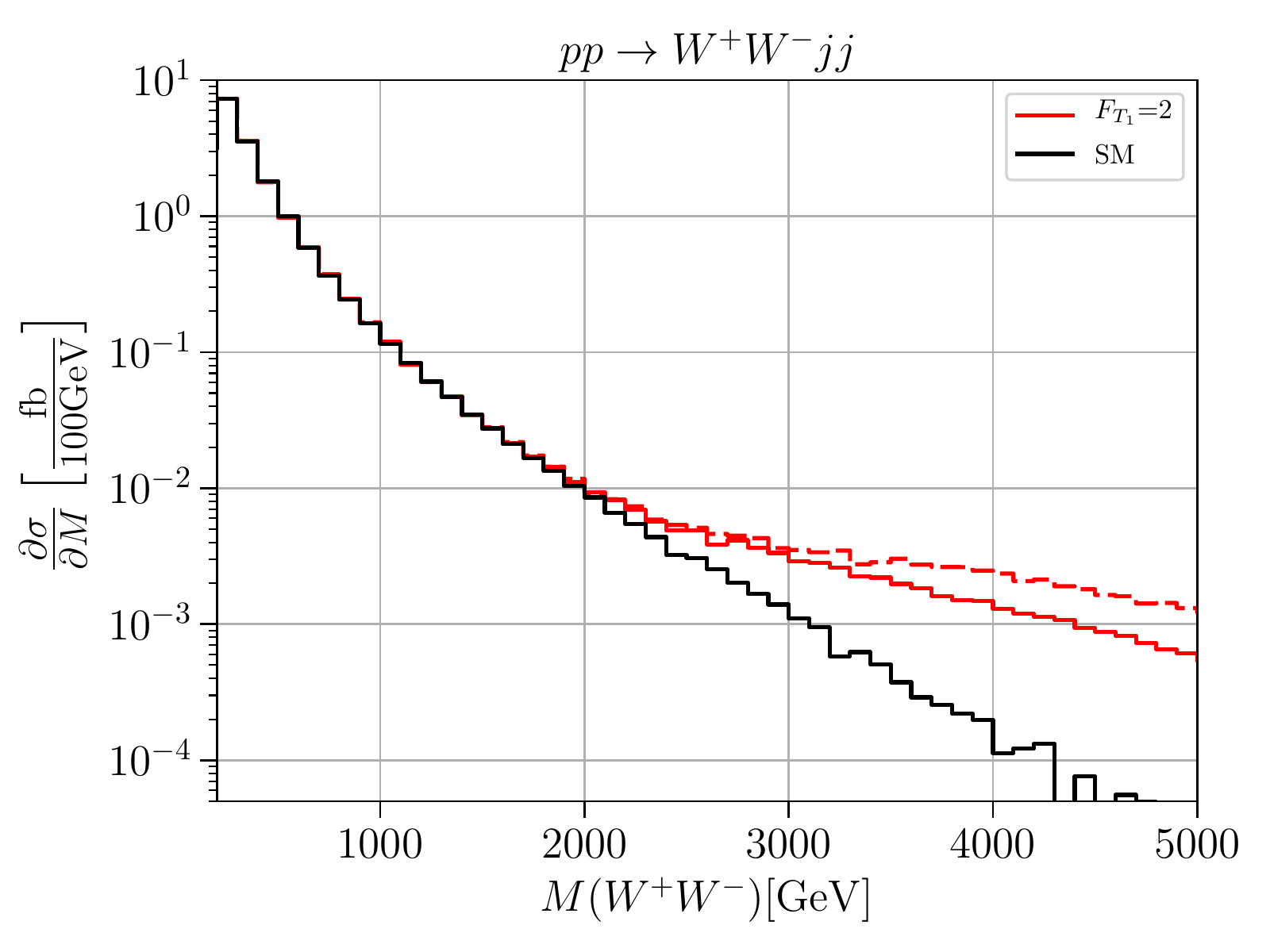}
    \caption{$\mathcal{O}_{T_1}$}
    \label{fig:WpWm-t1}
  \end{subfigure}\\[5pt]
  \begin{subfigure}{0.5\textwidth}
    \includegraphics[width=0.9 \linewidth]{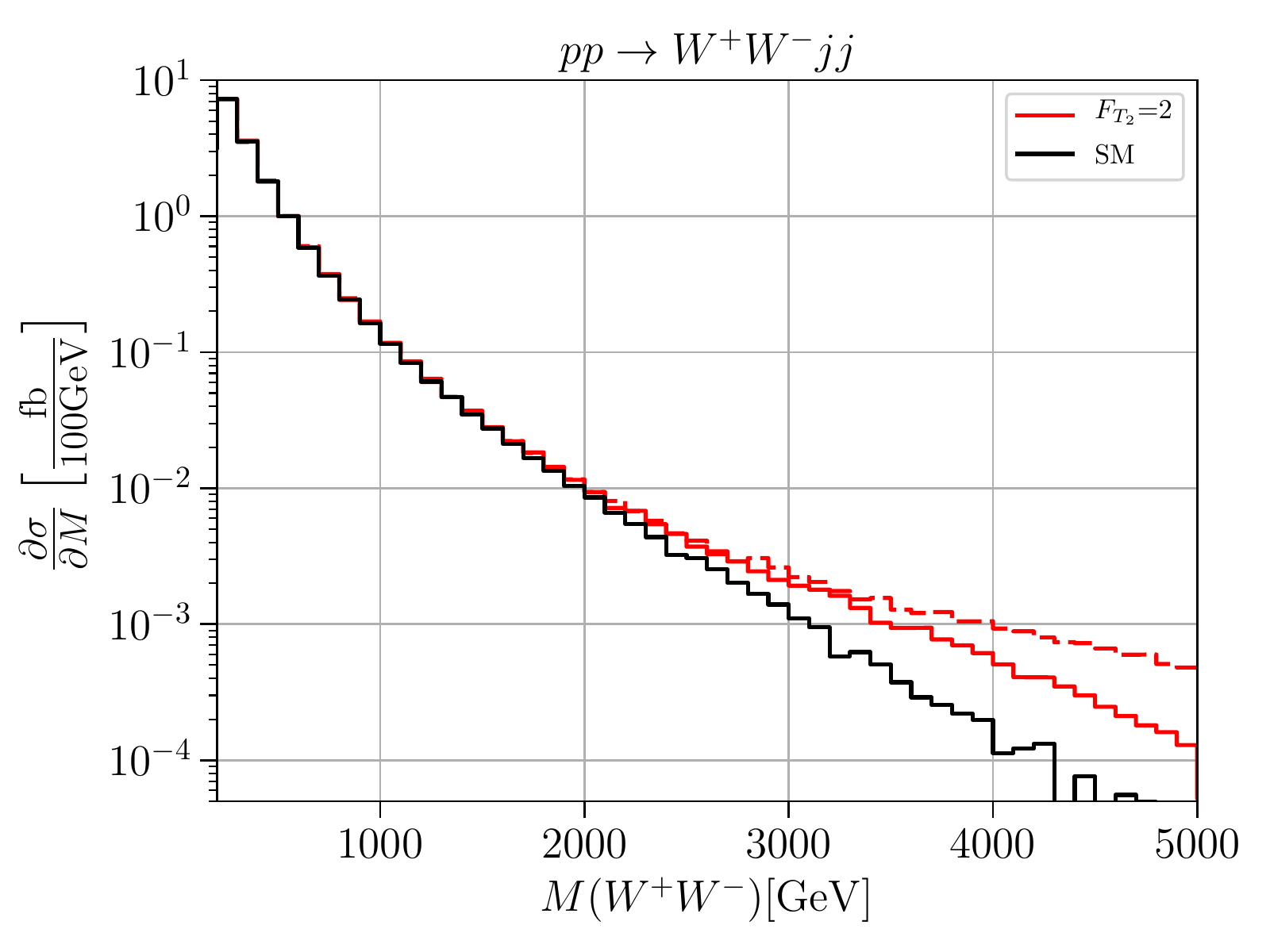}
    \caption{$\mathcal{O}_{T_2}$}
    \label{fig:WpWm-t2}
  \end{subfigure}
  \begin{subfigure}{0.5\textwidth}
    \includegraphics[width=0.9 \linewidth]{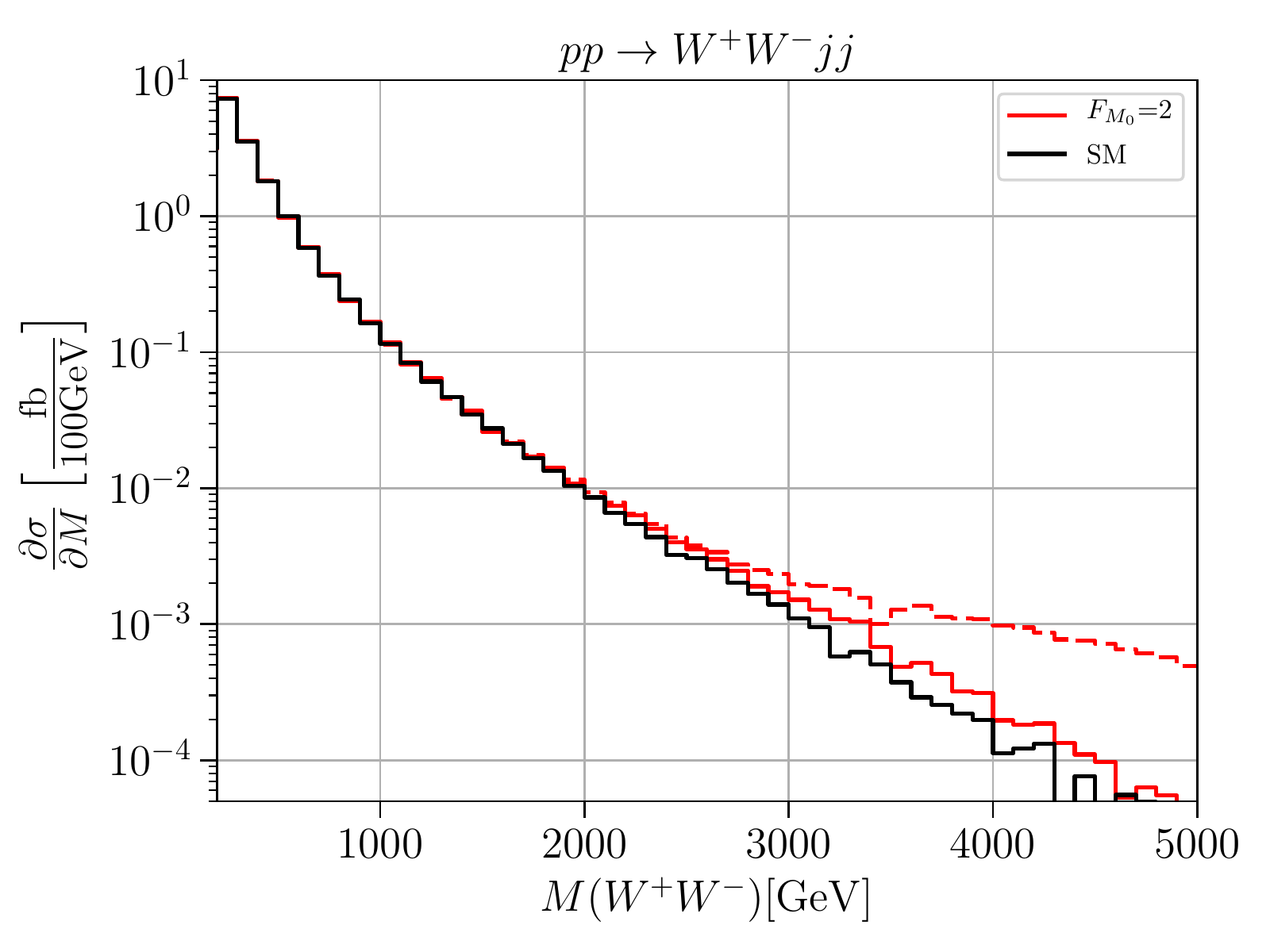}
    \caption{$\mathcal{O}_{M_0}$}
    \label{fig:WpWm-m0}
  \end{subfigure}\\[5pt]
  \begin{subfigure}{0.5\textwidth}
    \includegraphics[width=0.9 \linewidth]{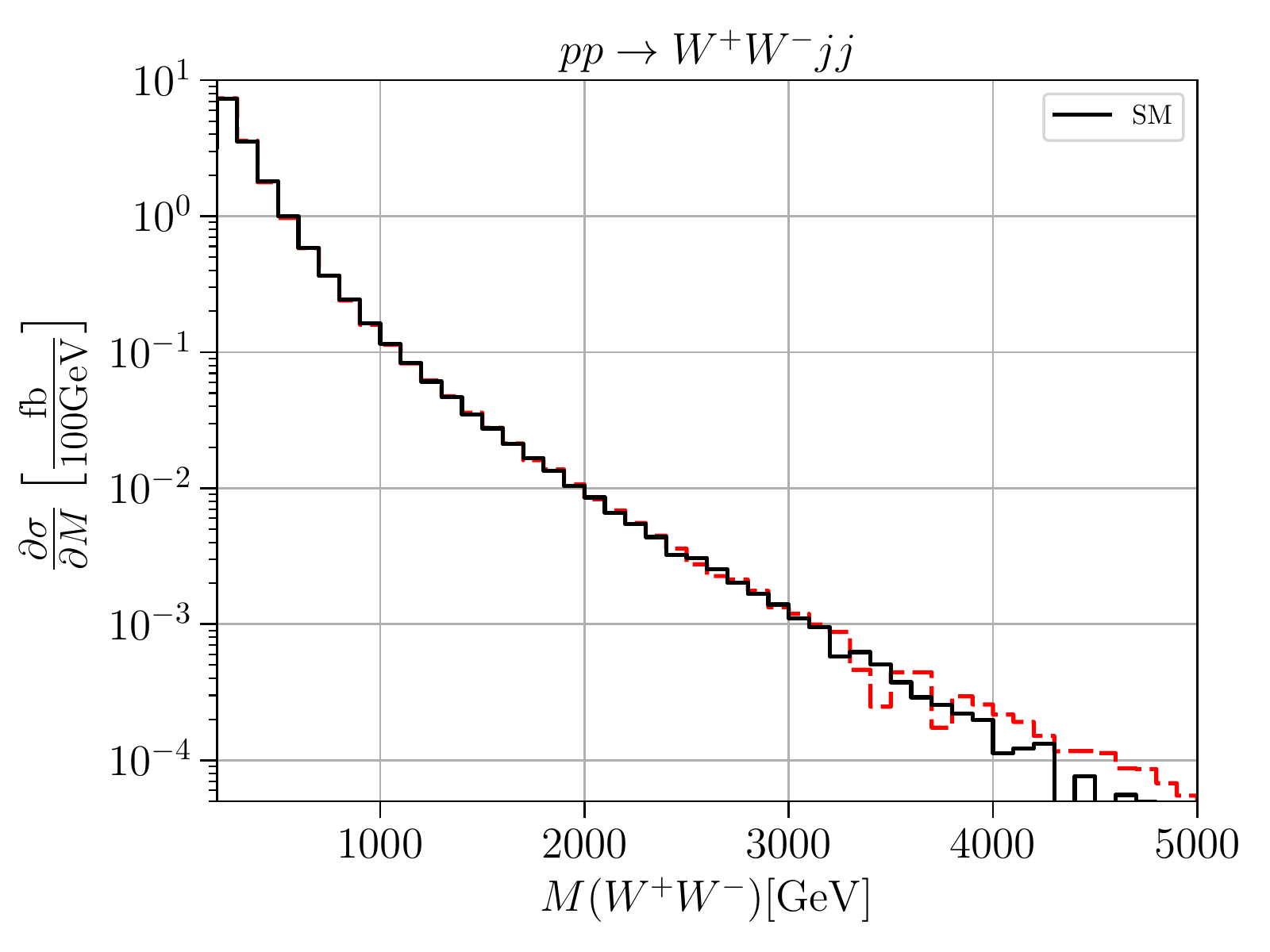}
    \caption{$\mathcal{O}_{M_1}$}
    \label{fig:WpWm-m1}
  \end{subfigure}
  \begin{subfigure}{0.5\textwidth}
    \includegraphics[width=0.9 \linewidth]{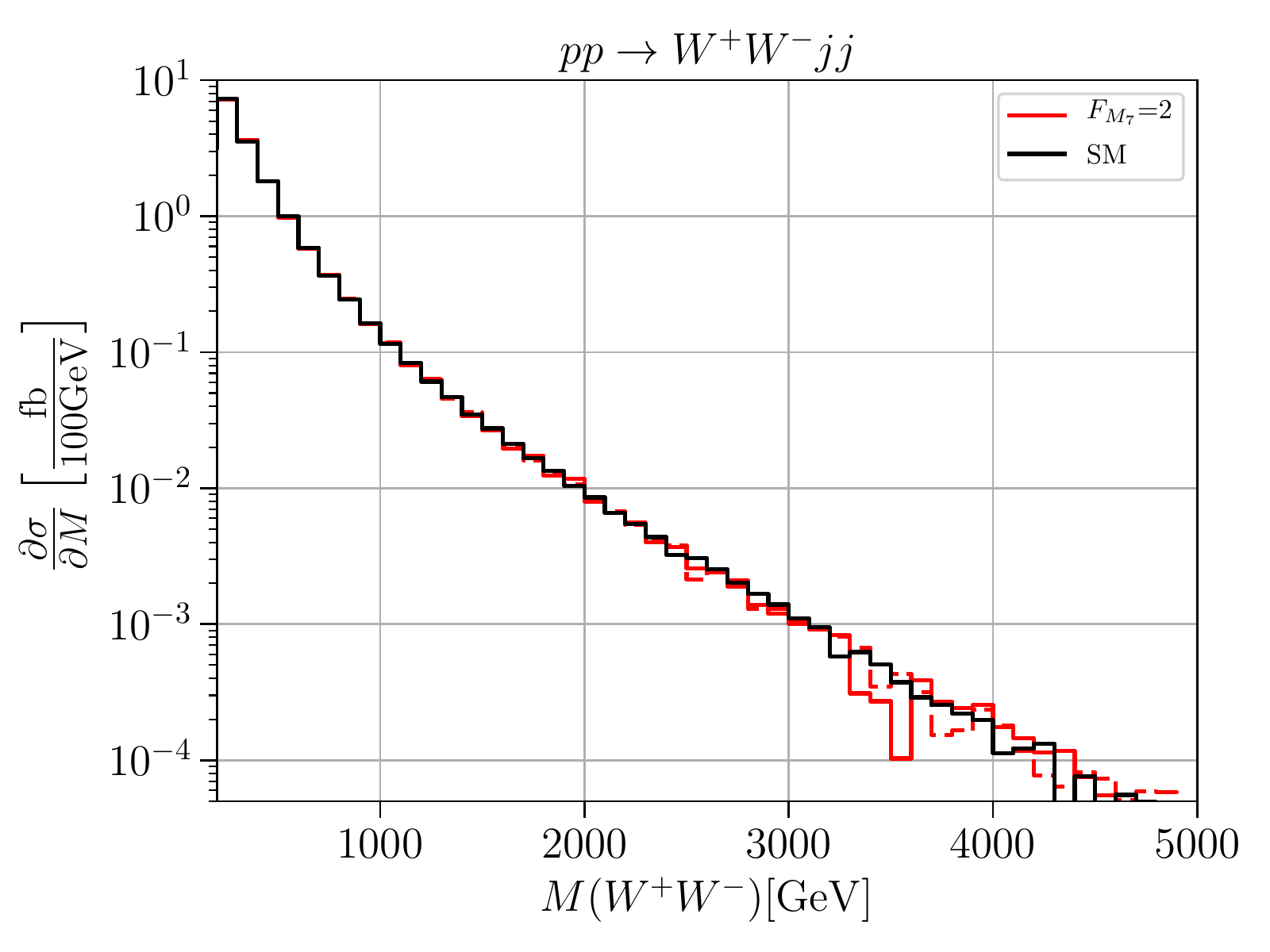}
    \caption{$\mathcal{O}_{M_7}$}
    \label{fig:WpWm-m7}
  \end{subfigure}
  \caption{The plot shows the differential cross section as a function of
    the invariant mass $m_{VV}$ of the two colliding vector bosons.
    The solid black line is
    the standard model and the colored lines are the
    contributions of $\mathcal{O}_{T_{0, 1, 2}}$ and $\mathcal{O}_{M_{{0, 1,
          7}}}$ (dashed: naive EFT, solid: unitarized model).
    Cuts: $M_{jj} > 500$ GeV;
    $\Delta\eta_{jj} > 2.4$;
    $p^j_T > 20$ GeV;
    $|\eta_j| > 4.5$
  }
  \label{fig:plot/WpWmjjft.pdf}
\end{figure}

\begin{figure}
  \begin{subfigure}{0.5\textwidth}
    \includegraphics[width=0.9\linewidth]{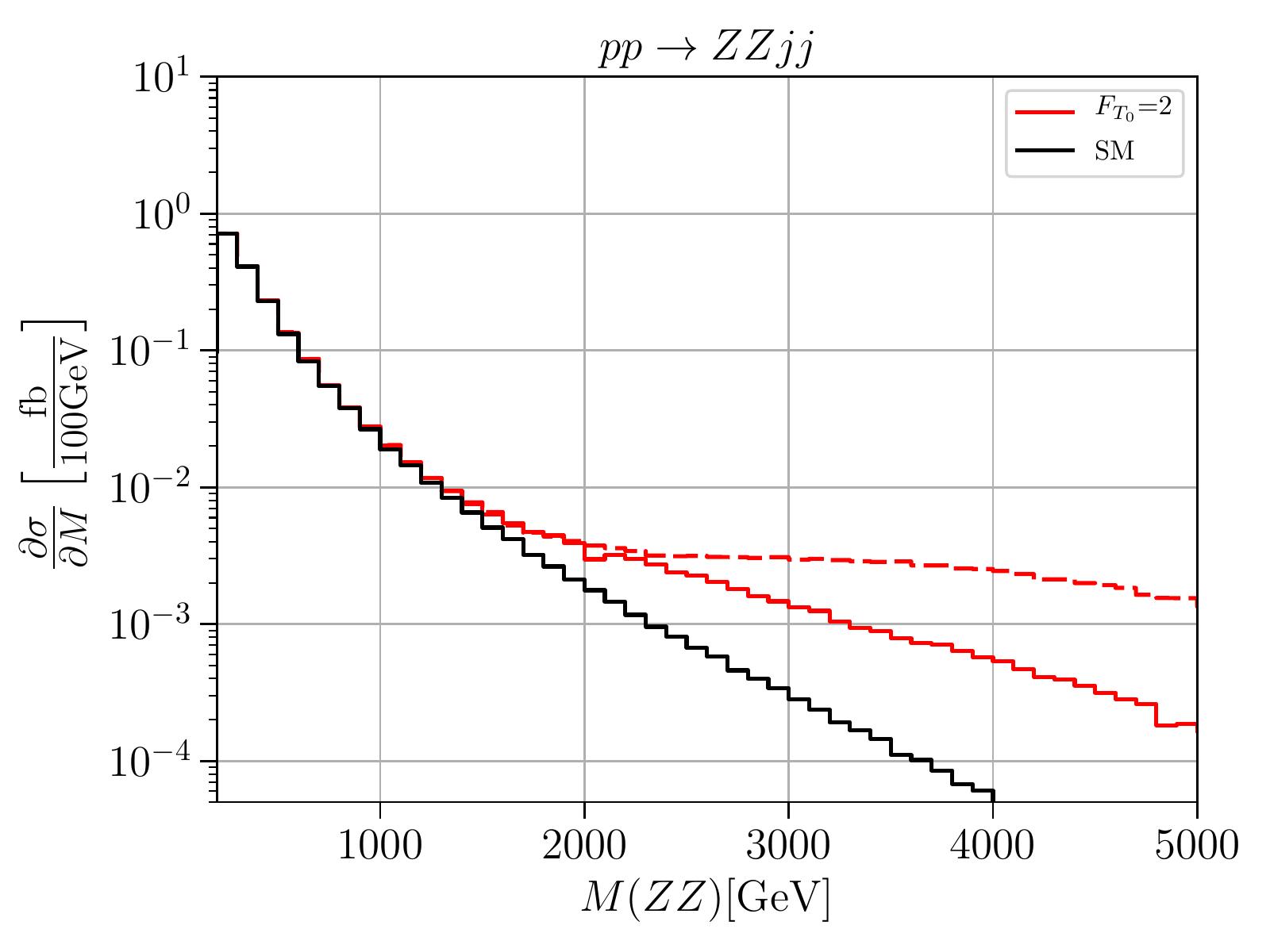}
    \caption{$\mathcal{O}_{T_0}$}
    \label{fig:ZZ-t0}
  \end{subfigure}
  \begin{subfigure}{0.5\textwidth}
    \includegraphics[width=0.9\linewidth]{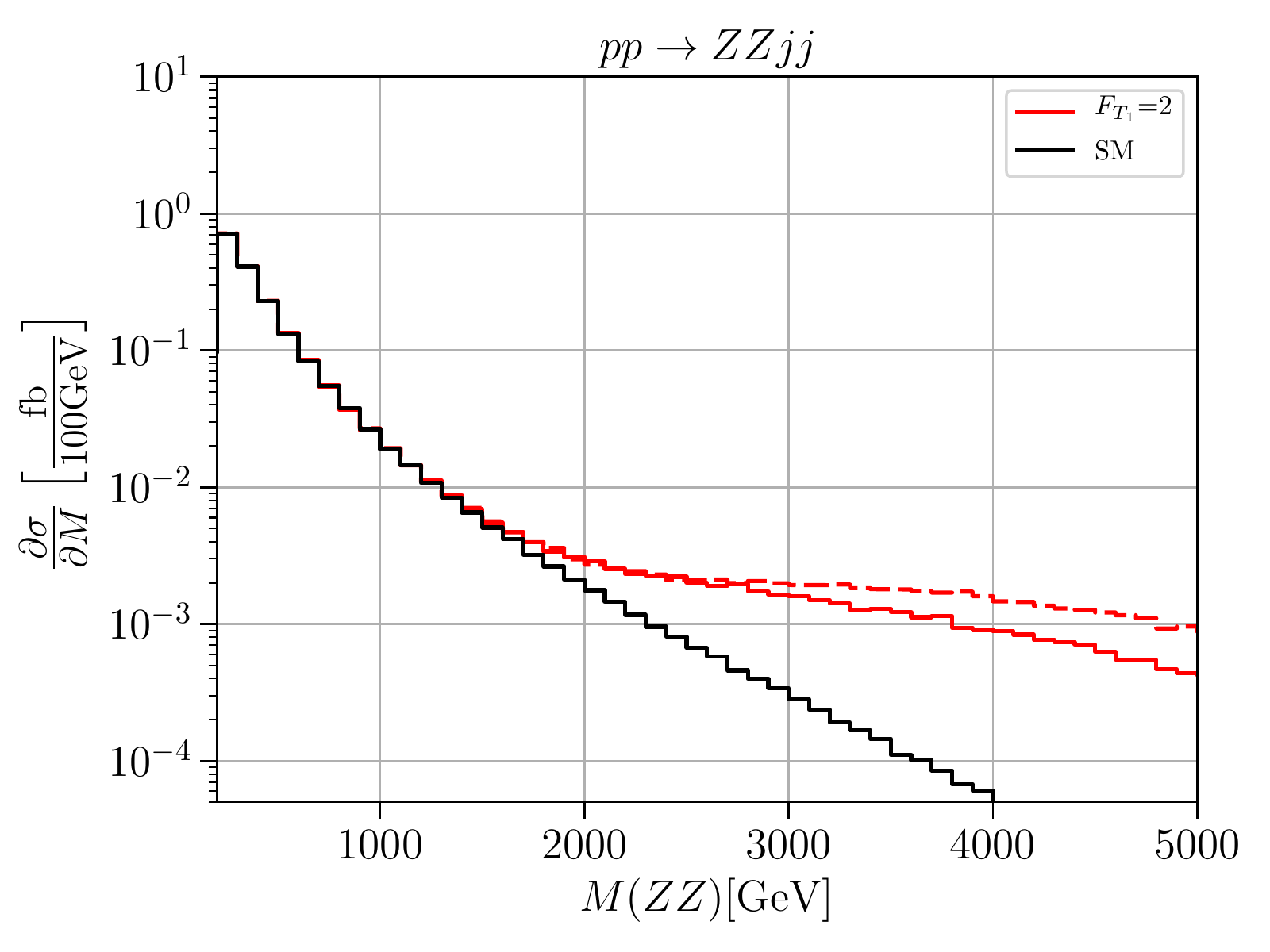}
    \caption{$\mathcal{O}_{T_1}$}
    \label{fig:ZZ-t1}
  \end{subfigure}\\[5pt]
  \begin{subfigure}{0.5\textwidth}
    \includegraphics[width=0.9\linewidth]{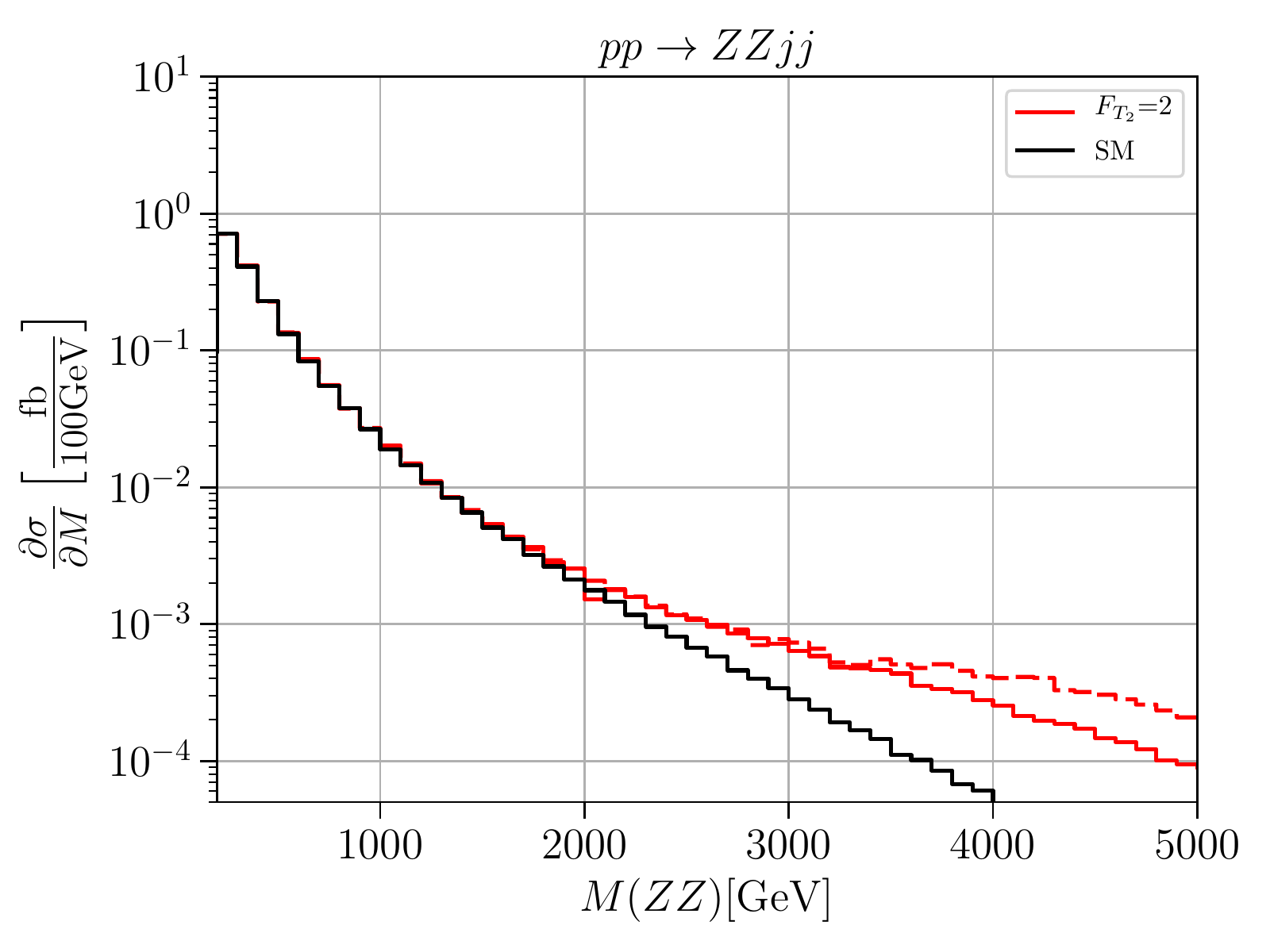}
    \caption{$\mathcal{O}_{T_2}$}
    \label{fig:ZZ-t2}
  \end{subfigure}
  \begin{subfigure}{0.5\textwidth}
    \includegraphics[width=0.9\linewidth]{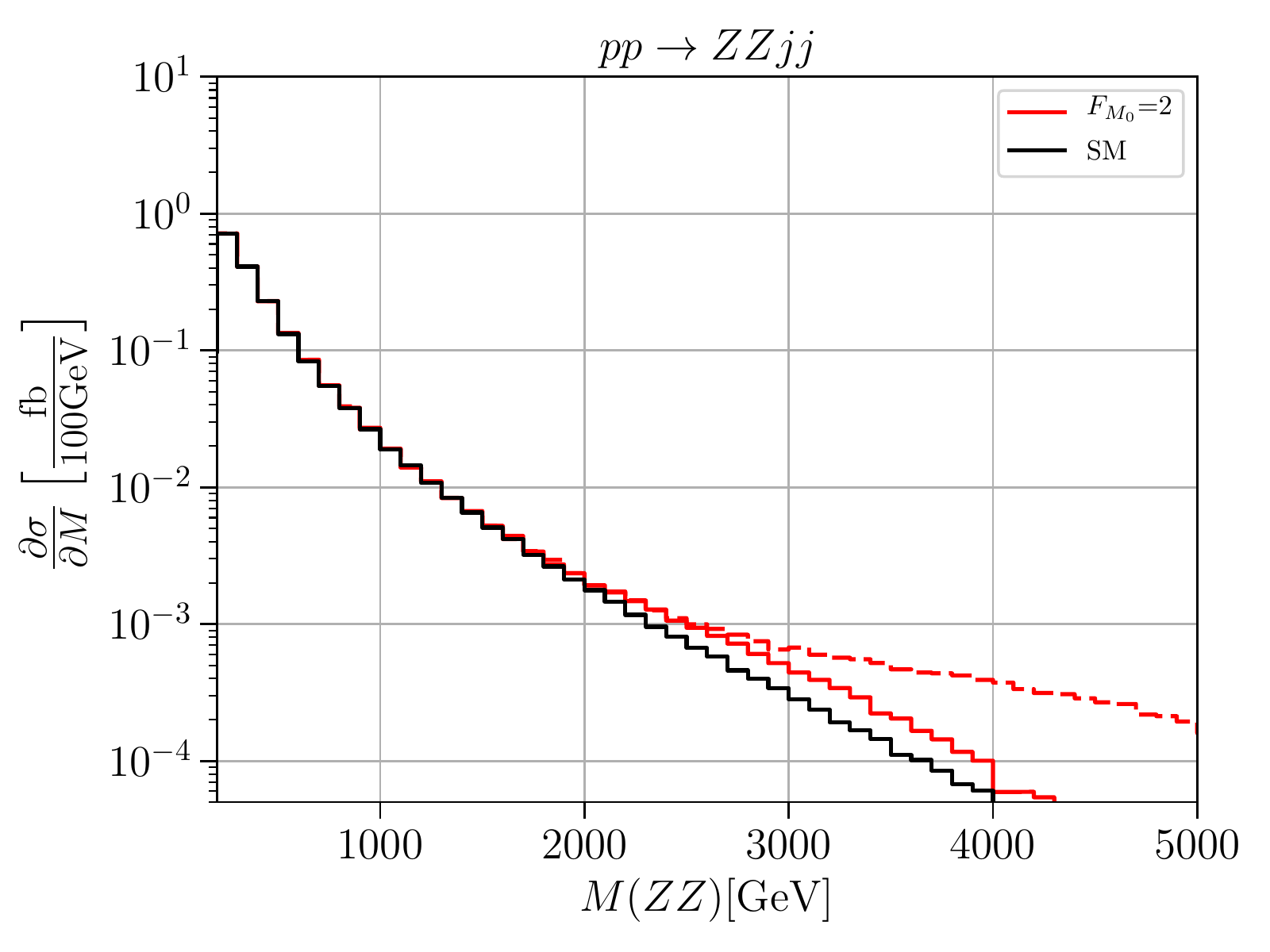}
    \caption{$\mathcal{O}_{M_0}$}
    \label{fig:ZZ-m0}
  \end{subfigure}\\[5pt]
  \begin{subfigure}{0.5\textwidth}
    \includegraphics[width=0.9\linewidth]{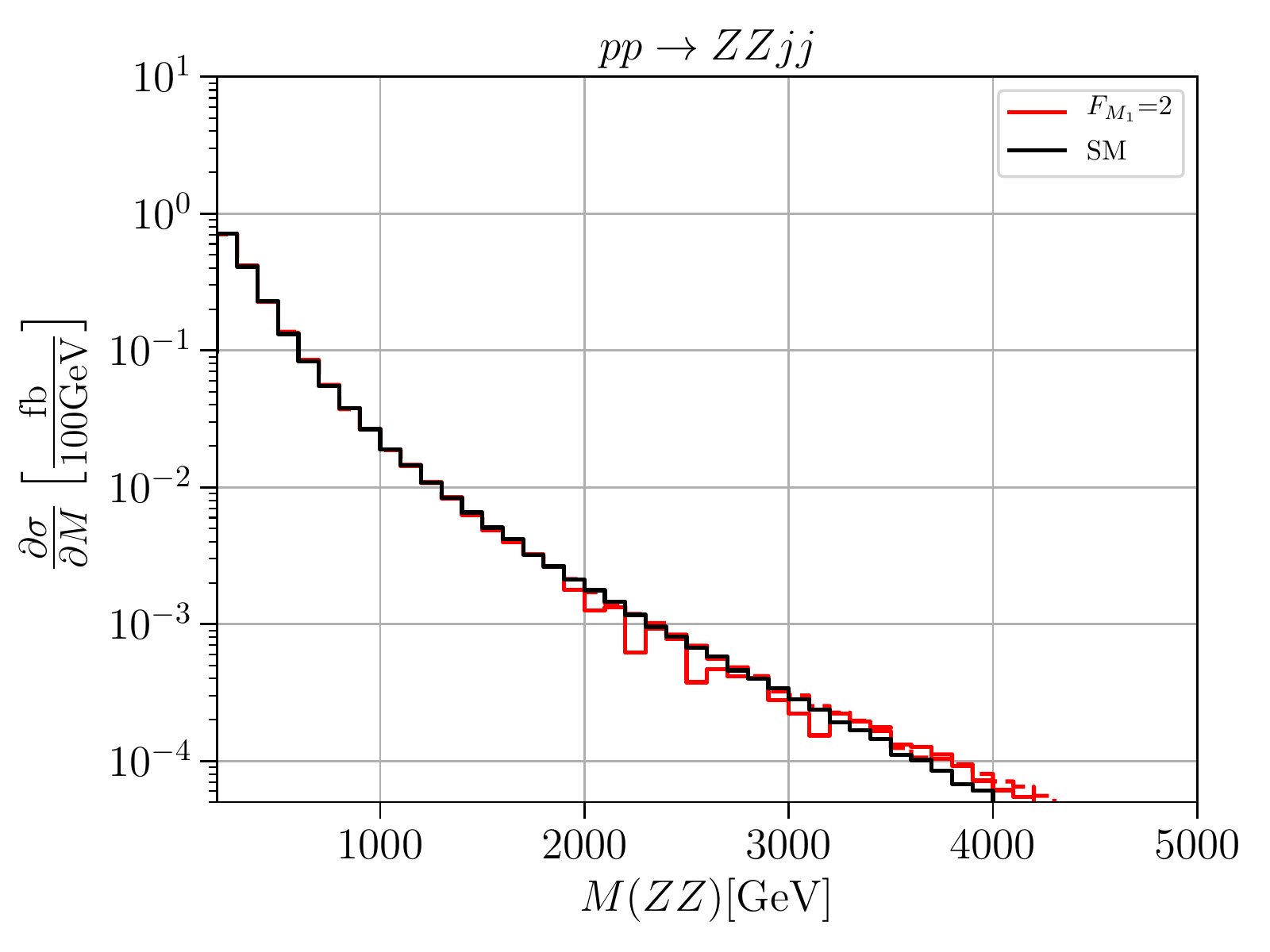}
    \caption{$\mathcal{O}_{M_1}$}
    \label{fig:ZZ-m1}
  \end{subfigure}
  \begin{subfigure}{0.5\textwidth}
    \includegraphics[width=0.9\linewidth]{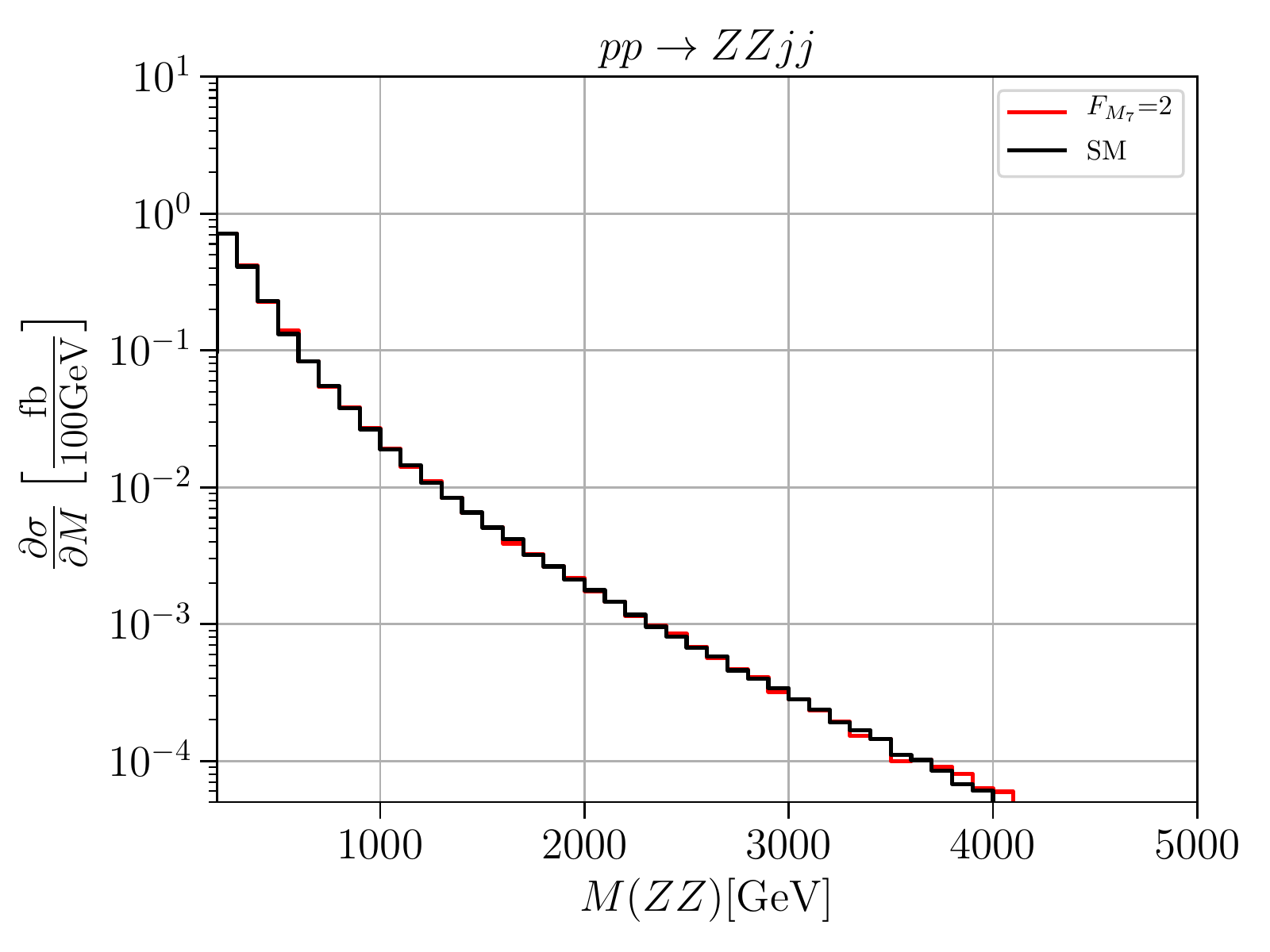}
    \caption{$\mathcal{O}_{M_7}$}
    \label{fig:ZZ-m7}
  \end{subfigure}
  \caption{The plot shows the differential cross section as a function of
    the invariant mass $m_{VV}$ of the two colliding vector bosons.
    The solid black line is
    the standard model and the colored lines are the
    contributions of $\mathcal{O}_{T_{0, 1, 2}}$ and
    $\mathcal{O}_{M_{0,1,7}}$ (dashed: naive EFT, solid: unitarized model).
    Cuts: $M_{jj} > 500$ GeV;
    $\Delta\eta_{jj} > 2.4$;
    $p^j_T > 20$ GeV;
    $|\eta_j| > 4.5$
  }
  \label{fig:plot/ZZjjft.pdf}
\end{figure}

\begin{figure}
  \begin{subfigure}[t]{0.5\textwidth}
    \includegraphics[width=0.9 \linewidth]{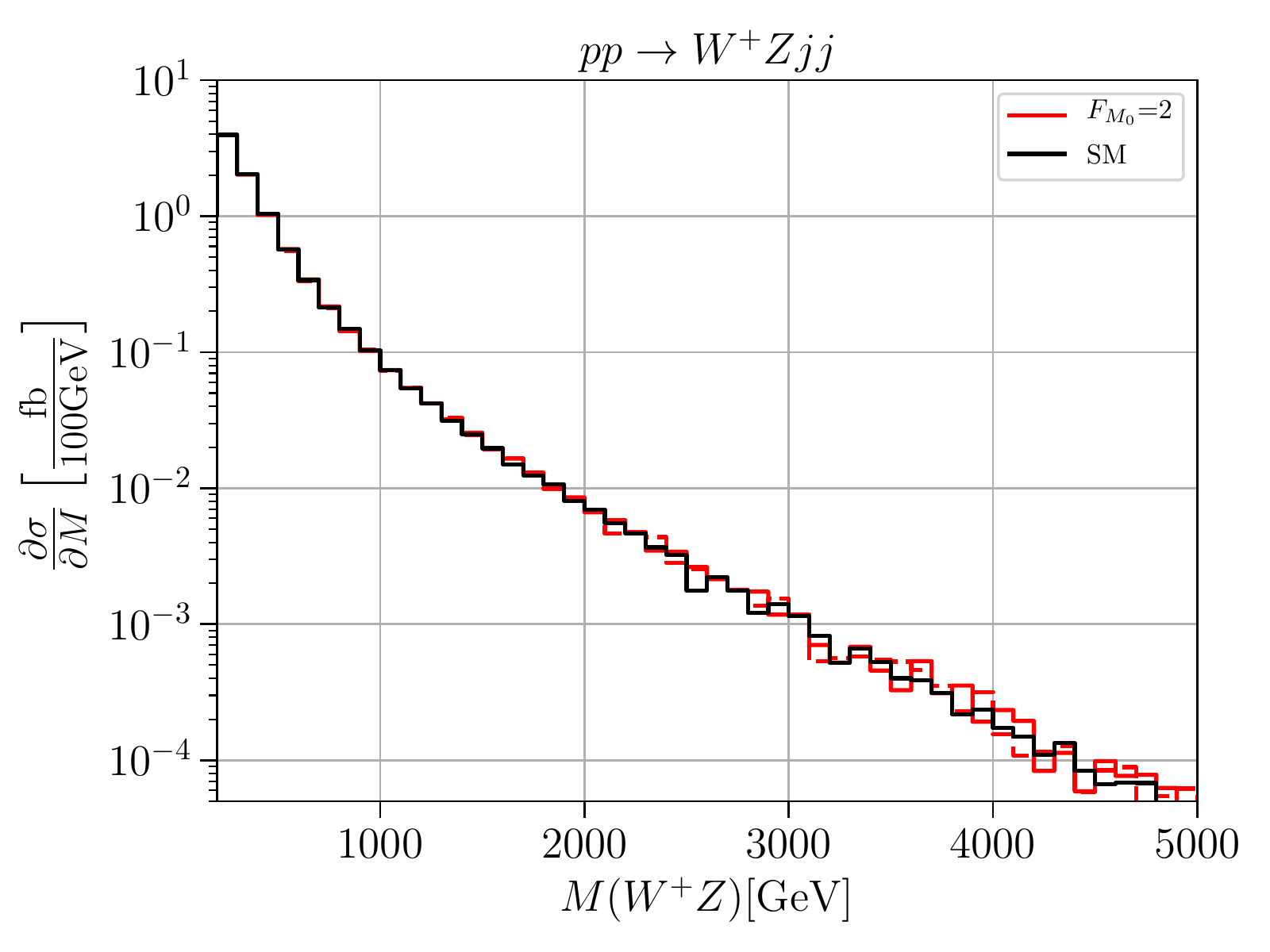}
    \caption{$\mathcal{O}_{M_0}$}
    \label{fig:WZ-m0}
  \end{subfigure}\hfill
  \begin{subfigure}[t]{0.5\textwidth}
    \includegraphics[width=0.9 \linewidth]{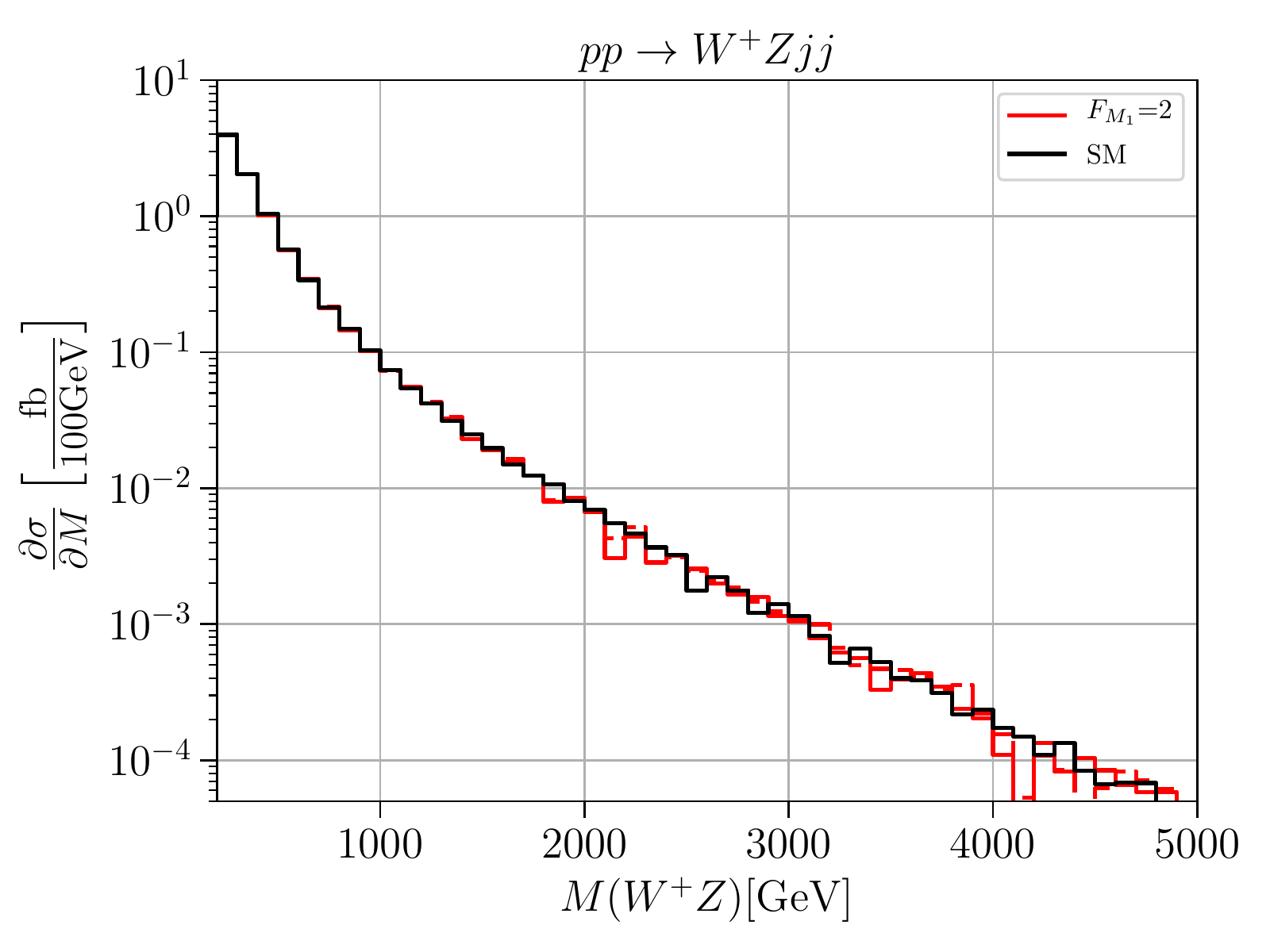}
    \caption{$\mathcal{O}_{M_1}$}
    \label{fig:WZ-m1}
  \end{subfigure}\\[5pt]
  \centering
  \begin{subfigure}[t]{0.5\textwidth}
    \includegraphics[width=0.9 \linewidth]{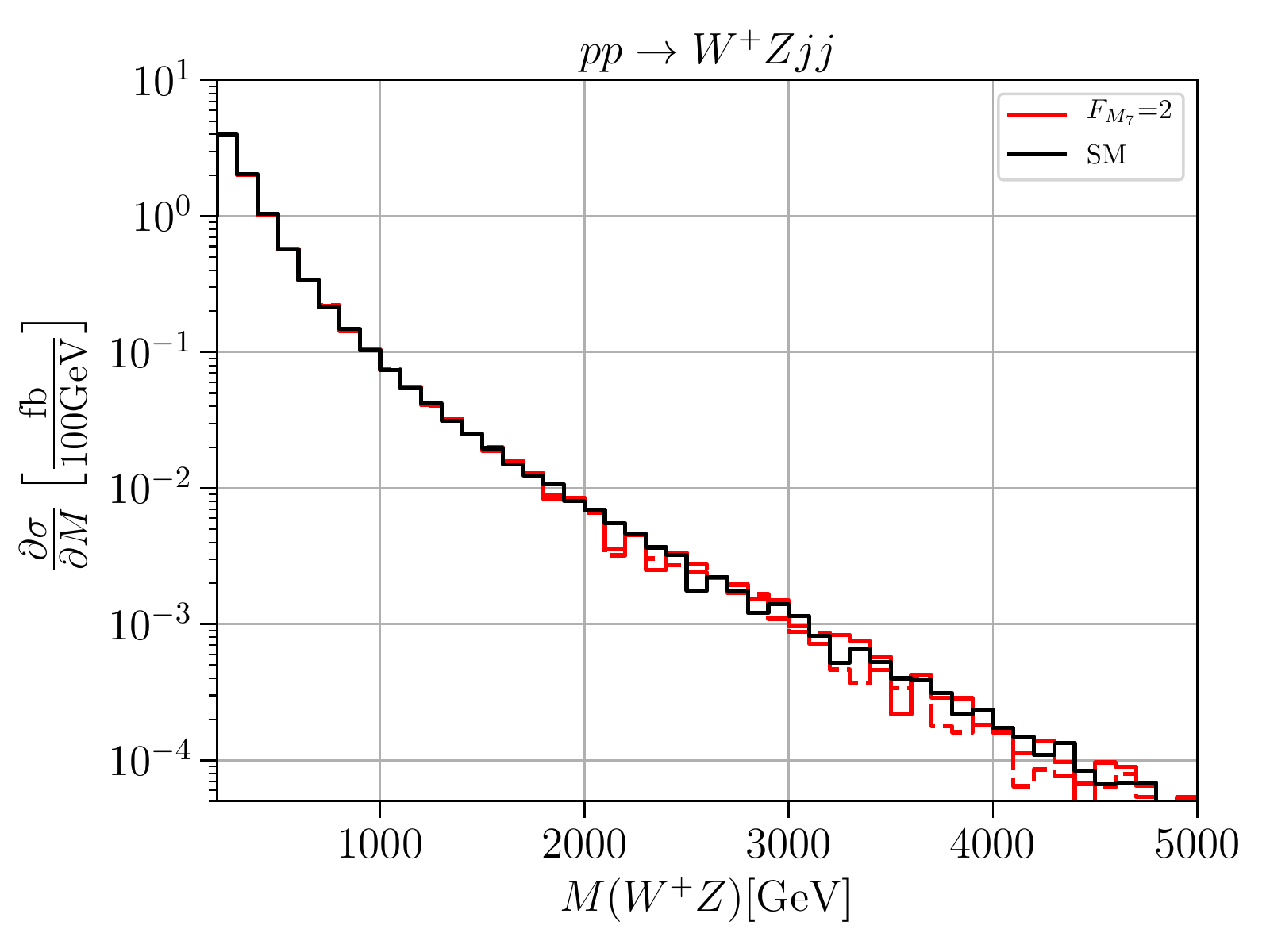}
    \caption{$\mathcal{O}_{M_7}$}
    \label{fig:WZ-m7}
  \end{subfigure}
  \caption{The plot shows the differential cross section as a function of
    the invariant mass $m_{VV}$ of the two colliding vector bosons.
    The solid black line is
    the standard model and the colored lines are the
    contributions of $\mathcal{O}_{M_{0,1,7}}$ (dashed: naive EFT, solid: unitarized model).
    Cuts: $M_{jj} > 500$ GeV;
    $\Delta\eta_{jj} > 2.4$;
    $p^j_T > 20$ GeV;
    $|\eta_j| > 4.5$
  }
  \label{fig:plot/WZjjfm0.pdf}
\end{figure}

\clearpage
\bibliographystyle{unsrt}

\end{document}